\documentclass[twoside, 3p, english]{elsarticle}
\usepackage{bm}
\usepackage[LGR,T1]{fontenc}
\usepackage{mathtools}
\usepackage{amsmath}
\usepackage{esint}
\usepackage{amsfonts}
\usepackage{color}
\usepackage{url}

\makeatletter

\ProvideTextCommand{\~}{LGR}[1]{\char126#1}

\newcommand{\lyxmathsym}[1]{\ifmmode\begingroup\def\b@ld{bold}
  \text{\ifx\math@version\b@ld\bfseries\fi#1}\endgroup\else#1\fi}

\makeatletter
\def\ps@pprintTitle{%
  \let\@oddhead\@empty
  \let\@evenhead\@empty
  \def\@oddfoot{\reset@font\hfil\thepage\hfil}
  \let\@evenfoot\@oddfoot
}
\makeatother
\biboptions{sort&compress}
\usepackage{geometry}
\begin{document}

\begin{frontmatter}{}

\title{\noindent Environment-friendly technologies with lead-free piezoelectric
materials: A review of recent developments, applications, and modelling approaches}

\author[rvt]{\noindent Akshayveer\corref{cor1}}

\ead{aakshayveer@wlu.ca}

\author[focal]{\noindent Federico C.~Buroni }

\author[rvt]{\noindent Roderick Melnik }

\author[els]{\noindent Luis Rodriguez-Tembleque }

\author[els]{\noindent Andres ~Saez }

\cortext[cor1]{\noindent Corresponding author}

\address[rvt]{\noindent MS2Discovery Interdisciplinary Research Institute, Wilfrid
Laurier University, Waterloo, Ontario N2L3C5, Canada }

\address[focal]{\noindent Department of Mechanical Engineering and Manufacturing,
Universidad de Sevilla, Camino de los Descubrimientos s/n, Seville
E-41092, Spain}

\address[els]{\noindent Department of Continuum Mechanics and Structural Analysis,
Universidad de Sevilla, Camino de los Descubrimientos s/n, Seville
E-41092, Spain}
\begin{abstract}
\noindent Piezoelectric materials are widely used in several industries, including power sources, energy harvesting, biomedical, electronics, haptic, photostrictive, and sensor/actuator technologies.
 Conventional piezoelectric materials, such lead zirconate titanate (PZT), pose significant environmental and health risks due to lead content.  Recent years have seen a growing need for eco-friendly alternatives to lead-based piezoelectric technologies.  The drawback of lead-free piezoelectric materials is a reduced responsiveness.  To enhance the performance of lead-free piezoelectric materials, substantial research is being undertaken using both experimental and numerical methods.  The experimental studies provide a deep understanding of the process and are crucial for developing lead-free piezoelectric materials.  Relying only on experimental research is not feasible due to high expenditures.
 To understand the piezoelectric properties of lead-free materials and enhance their efficiency, it is crucial to develop varied numerical models and computational methods.  This paper is a comprehensive summary of lead-free piezoelectric technology breakthroughs.  The study emphasizes numerical models that demonstrate enhanced piezoelectric behaviour and their use in eco-friendly technologies like energy harvesting, haptics, nano-electromechanical, photostrictive, and biomedical sensors and actuators using lead-free materials.  The paper discusses the synthesis, properties, and applications of eco-friendly materials, highlighting their potential to revolutionize piezoelectric devices and promote sustainable development and conservation.
 
\end{abstract}

\end{frontmatter}{\noindent }

\noindent 

\section{Introduction}
\markright{Introduction \hfill \empty}

As the world continues to place a premium on sustainability and environmental
awareness, this review paper aims to shed light on promising advancements
in lead-free piezoelectric technologies. The transition to eco-friendly
piezoelectric materials can help mitigate environmental risks while
promoting technological progress and innovation by raising awareness
and encouraging research in this field. Precision actuators, energy
harvesting devices, high-frequency sensors, and ultrasonic transducers
are just a few of the applications that can be enabled by piezoelectric
technologies. The ability of piezoelectric materials to convert mechanical energy to electrical energy and vice versa has revolutionized many
technological fields \citep{IbnMohammed2017}. Lead-based compounds,
particularly lead zirconate titanate (PZT), have dominated the piezoelectric
material market due to their exceptional performance and reliability
\citep{Krishnaswamy2020,Krishnaswamy2019,kvasov2016}.

Despite their success, lead-based piezoelectric materials have raised
significant environmental and health concerns \citep{Panda2009}.
Lead is a toxic heavy metal that has been linked to serious health
problems, particularly in children and vulnerable populations, even
at low levels of exposure \citep{Collin2022}. Furthermore, improper
disposal or recycling of lead-containing devices can cause soil and
water contamination, endangering ecosystems and human health. Researchers
and industries have been increasingly motivated to investigate viable
alternatives to lead-based piezoelectric materials in response to
the pressing need for environmentally sustainable technologies \citep{Wu2003}.
\citep{Li2018,Liu2018}.

\noindent Lead-free piezoelectric composites provide scalable and
environmentally sustainable choices for electromechanical actuators
as well as mechanical energy harvesting and sensing \citep{Maurya2018,Tiller2019,IbnMohammed2017,zhao2019}.
Lead-free piezocomposites that can be printed in three dimensions (3D) provide scalable and ecologically friendly solutions for many technical applications. Usually, the composite system is made up of polymeric matrices that are strengthened with active polycrystalline particles and maybe nanoadditives. Interfacial inclusion/matrix degradation not only undermines the structural integrity of the component but also profoundly affects its functionality as smart materials \citep{Canamero2023,Canamero2024}. These composites typically consist of softer matrices, usually polymers,
filled with small particles of hard, crystalline, lead-free piezoelectric
materials that exhibit strong piezoelectric activity. Examples of
these materials include non-perovskites like bismuth layered structured
ferroelectrics (BLSF) and tungsten bronze ferroelectrics \citep{Takenaka2005},
lithium-based ferroelectric materials \citep{Zhang2015}, and perovskites
like BaTiO$_{3}$ (BT), Bi$_{0.5}$Na$_{0.5}$TiO$_{3}$ (BNT), and K$_{0.5}$Na$_{0.5}$NbO$_{3}$
(KNN) \citep{Shin2014,Quan2021,Zhang2023}, etc. Lead-free piezoelectric
materials have reduced piezoelectric coefficients in their most pure
state. However, their piezoelectric performance can be improved through
inherent factors such as lattice distortions and exogenous factors
such as modifying the domain topologies \citep{Damjanovic1998}. Moreover,
certain impurities or secondary materials are mixed in these lead-free
materials to enhance their electromechanical properties. A novel piezoelectric
material based on potassium sodium niobate (KNN) demonstrates enhanced
piezoelectric response (d$_{33}\sim$ 500 pc/N) and a higher Curie temperature
(T$_{c}\sim$ 200$\lyxmathsym{\textcelsius}$) by reducing lattice softening
and minimizing unit cell distortion \citep{Liu2019}. In $(1-x)$BaTiO$_{3}$-xCrZnO$_{3}$,
a remarkably high piezoelectric response of (d$_{33}\sim$ 445$\pm$20 pc/N)
is produced by controlling the multiphase coexistence of rhombohedral-orthorhombic-tetragonal
(R-O-T) and the phase transition temperature \citep{Wang2021}. This
is also a perfect example of the addition of secondary elements such as
CrZnO$_{3}$ in pure lead-free BaTiO$_{3}$, which alters the domain
topology of the pure BT material and enhancement of piezoelectric response
is seen. Similary, the piezoelectric properties of pure BNT (which
has depicted moderate piezoelectric properties (d$_{33}\sim$ 100 pc/N)
and high coercive electric field (E$_{c}\sim$ 70 kV/cm) \citep{Hao2019})
with certain amounts of BT doping have been enhanced by engineering
template grain formation utilizing NN templates, which harnessed the
microstructure of BNT-based ceramics \citep{Bai2018}.

\noindent The aforementioned literature clearly indicates that there
is significant potential for advancement and innovation in the field
of lead-free piezoelectric materials and the energy harvesting systems
that rely on them. In recent years, numerous experimental research
works \citep{Yin2018,Liu2015,Hou2022,Pattipaka2018,Jarupoom2022}
have been done to enhance the piezoelectric performance of lead-free
piezoelectric materials. These experimental studies yield valuable
insights into the piezoelectric properties of lead-free materials.
However, the necessity for repeated studies to enhance comprehension
renders these experiments costly and time-consuming, with significant
wastage of the piezoelectric material during the examination process.
Consequently, many researchers \citep{Krishnaswamy2020a,Krishnaswamy2019a,Yang2006,Chaterjee2018,Krishnaswamy2020b}
have devised diverse numerical models to extensively investigate piezoelectric
materials. The aim is to derive new theories from these analyses and
obtain more efficient versions of lead-free piezoelectric materials
by utilizing their findings in a cost-effective and time-efficient
manner. This review paper aims to provide an in-depth examination
of environmentally friendly piezoelectric technologies based on the
use of lead-free materials. Our primary focus is on discussing recent
advancements, with a particular emphasis on the potential applications,
and various numerical models developed for synthesis of these eco-friendly
materials and analysis of their properties. We seek to assess the
viability of lead-free piezoelectric technologies and their significance
in achieving a more sustainable and greener future by considering
environmental implications, life cycle analysis, and performance characteristics.

The rest of the paper is organized as follows: Section 2 provides
an overview of lead-free piezoelectric materials, providing details
of most-promising material classes, recent advancements, and the performance
of the piezoelectric materials. Section 3 gives a comprehensive overview
of abundant applications of the piezoelectric materials, emphasizing
the importance of lead-free piezoelectric materials in these fields
of applications. Section 4 focuses on the main research directions
in developing the numerical models to analyze properties of these
materials and providing enhancive measures of the electromechanical
performance of piezoelectric-based applications. Finally, in Section
5, we summarize the importance of modelling advances made in lead-free
piezoelectric technologies, emphasizing their environmental benefits
and long-term implications, and provide our perspective on future directions
in these areas. We emphasize the importance of continuing R\&D efforts
to accelerate the adoption of these materials in mainstream industries.

This review paper aims to contribute to the growing body of knowledge
on environmentally responsible piezoelectric technologies. We hope
to inspire further exploration and innovation in the pursuit of greener
and more efficient piezoelectric devices by fostering a deeper understanding
of lead-free materials and their potential applications.
\begin{figure}[t]
    \centering
    \includegraphics[width=0.75\linewidth]{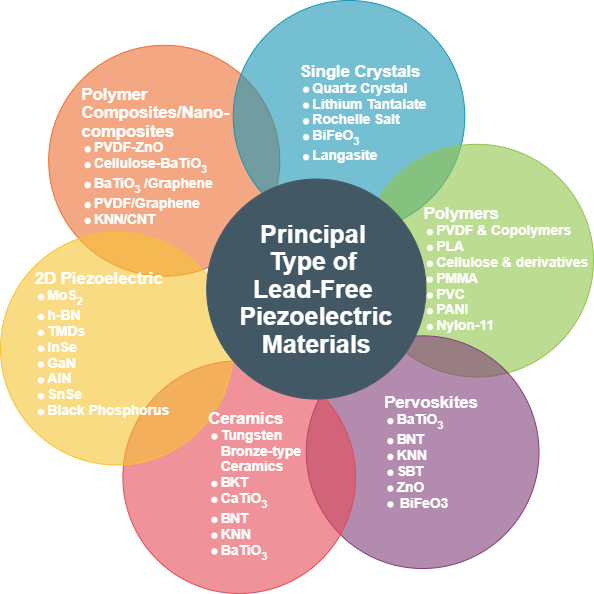}
    \caption{Schematic diagram of principal types of piezoelectric materials}
    \label{fig:Schematic1}
\end{figure}
%
\section{Background: Principal types of lead-free piezoelectric materials}
\markright{Background: Principal types of lead-free piezoelectric materials \hfill \empty}
Lead-based piezoelectric materials, particularly PZTs, exhibit significantly
good piezoelectric response and dominate the whole piezoelectric's market.
However, the lead is a toxic material itself, and it has various hazardous
effects on human health and the environment. Therefore, a lot of research
has been carried out on potential lead-free alternatives of PZTs such
as BNT, BT, KNN, BLSF, and lithium-based piezoelectrics, as all of
these options suffer from the low piezoelectric response. The lead-free
piezoelectric can also be a favourable option for cochlear implants,
as the disadvantage of harmful lead has been avoided.

Pure BT and KNN ceramics have a substantially lower piezoelectric
coefficient (d$_{33}\sim$ 200 pC/N) than PZT-based ceramics. Modified
BT-, KNN-, and BFO-based ceramics, on the other hand, have comparable
d$_{33}$ values to PZT-based ceramics. The use of phase boundary
engineering (PBE), in particular, has resulted in non-textured BT-
and KNN-based ceramics with d$_{33}$ values as high as 490-700 pC/N
\citep{Liu2009,Wang2014,Wu2016}, and when combined with the RTGG
technique, their d$_{33}$ values grew to more than 700 pC/N. 
Various technologies, such as phase boundary engineering, optimizing
composite design, composite ceramics, sintering, and preparation technology,
can have inherent and extrinsic impacts on piezoelectrics, enhancing
their piezoelectric response. Both intrinsic and extrinsic contributions
\citep{Damjanovic1998,Li2018a,Liu2018a} are considered when describing
the piezo/ferroelectric characteristics of lead-free piezoceramics.
The intrinsic gift is connected to their lattice arrangement (for
example, lattice distortion). In contrast, the extrinsic contribution
is mostly determined by their microstructure, which includes ferroelectric
domains, particle size, density, and porosity. However, most acoustic
sensors based on lead-free piezoelectric materials have failed to
provide high sensitivity, desired flexibility, broad frequency selectivity,
and biocompatibility simultaneously. The limitations have been removed
by introducing a micro-cone patterning array strategy based on novel
lead-free, multicomponent rod-shaped niobate piezoelectric materials
with a morph-tropic phase boundary. The rod-based, flexible piezoelectric
acoustic sensor's (FPAS) output voltage is almost three times as high
as the isotropic particles. Due to the large improvement in sound
energy absorption, the micro-cone patterning FPAS with consistent
performance in hostile conditions exhibits high sensitivity (39.22
mVPa$^{-1}$cm$^{-2}$) and can record sound signals, identify audio
signals, and realize human-computer interactions \citep{Xiang2023}.
Because the piezoelectric response of piezoelectric ceramics only
appears after poling engineering, the domain evolution and microscopic
piezo-response were observed in situ using piezo-response force microscopy
(PFM) and switching spectroscopy piezo-response force microscopy (SS-PFM),
which can effectively study the local switching characteristics of
ferroelectric materials, particularly at the nanoscale, in this work.
The new domain nucleation develops preferentially near the edge of
the relative polarization zone and extends laterally as bias voltage
and temperature rise. The comparable d$_{33}$ at 50 kV/cm and
120$\lyxmathsym{\textcelsius}$ achieves a maximum of 205 pC/N,
approximately twice as high as that at room temperature, due to the
specific influence of poling engineering \citep{Song2023}. 

Current high-performing piezoelectric materials, particularly those used in advanced applications, are predominantly dominated by perovskite structures. These materials rely on soft optical phonon modes, which are stabilized by disorder near a morphotropic phase boundary (MPB) \cite{bouchbinder2018universal}. This MPB not only influences the material's lattice dynamics but also plays a crucial role in the unique resilience of the polar response to such disorder \cite{hirata2019first}. The ability of these materials to maintain a strong piezoelectric response in the presence of structural disorder has made them highly effective, yet understanding the underlying mechanisms remains complex.
To explore this phenomenon further, researchers have developed a first-principles sensitivity analysis approach aimed at assessing the effects of disorder on the piezoelectric response of materials \cite{baroni1987greens}. This approach leverages data from the Materials Project database, which is a valuable resource for identifying potential candidate materials based on their structural properties and predicted behavior. The analysis reveals that in many well-known piezoelectric systems, the lattice dynamics—rather than internal strain or dielectric properties—play a more significant role in controlling the polar response \cite{kormann2017phonon}. This insight shifts the focus toward understanding the phonon modes within these materials, which are instrumental in determining their piezoelectric properties.

A key finding from the study is the discovery of multiple stable optical phonon modes that contribute to the piezoelectric response. These modes act as a kind of "fingerprint" for the material’s tolerance to disorder, offering a way to predict how a material will perform when exposed to structural imperfections \cite{zhang2007modified}. This observation has led to the formulation of a multiple-phonon mode criterion, which can be applied to evaluate and select candidate materials for new piezoelectric prototypes that are more tolerant of disorder. This criterion provides a systematic method for discovering novel materials with superior piezoelectric properties, particularly those that go beyond the capabilities of traditional perovskite-based systems.
Through the use of this criterion, researchers identified five promising materials that, when altered through chemical substitution, demonstrate potential as new piezoelectric prototypes \cite{zheng2015effects}. These materials exhibit large piezoelectric responses that rival or exceed those of perovskites. One such material, Akermanite Sr$_{2x}$Ca$_{2-2x}$CoSi$_2$O$_7$, stands out due to its significant improvement in piezoelectric performance. At a 50\% composition, this material shows nearly a 20\% increase in response, indicating its potential to be a valuable alternative to conventional piezoelectric materials, offering enhanced performance in applications that require high sensitivity and durability \citep{Ling20207}. This approach, by identifying and optimizing materials with disorder-tolerant piezoelectric responses, opens up new avenues for the development of next-generation piezoelectric devices with superior functionality and stability.

In this
section, a brief has been provided on these lead-free alternatives
to PZT and other lead-based piezoelectric materials. The piezoelectric response
of lead-free piezoelectric materials, such as KNN, BNT, BT, and Li-based
materials, is  relatively low compared to conventional lead-based materials such as PZT. Hence, this section provides an overview of the extensive literature on enhancing these materials' piezoelectric performance through numerous innovative techniques such as phase boundary engineering (PBE) and creative multiscale designs of piezoelectric composites.

\subsection{Potassium sodium-niobate (KNN)-based piezoelectrics}

It has been observed that modifying the polarization rotation of potassium
sodium-niobate (KNN) lead-free piezoelectric materials improves piezoelectric
properties and piezo-catalytic activities. The effectiveness of modulating
polarization rotation was demonstrated by degrading Rhodamine B (RhB)
using three KNN-based samples with different phase structures (i.e.,
orthorhombic (O) phase, orthorhombic-tetragonal (O-T) coexistence
phase, and rhombohedral-orthorhombic-tetragonal (R-O-T) coexistence
phase) \cite{lin2024,zheng2022}. The response rate constant of poled samples with the R-O-T
coexistence phase is 0.091 min$^{-1}$, owing to the most effortless
polarization rotation, which is 2.12 times bigger than the reaction
rate constant of poled O-phase samples with the most challenging polarization
rotation. Improved piezo-catalytic activities are primarily due to
easier polarization rotation and enhanced carrier concentration, followed
by mechano-charge generation \citep{Sun2022}. Transparent piezoelectric
ceramics have sparked considerable interest in recent years due to
their electro-optical uses. However, due to the inherent trade-off,
it isn't easy to obtain both strong piezoelectricity and transparency
simultaneously, particularly in lead-free ceramics. Sm-doped KNN-based
ceramics exhibit a higher openness of 63\% and a higher piezoelectric
response of 153 pC/N at a concentration of $x=0.004$. The presence
of a pseudo-cubic phase and the significant elimination of domain
walls by the Sm modification are responsible for the high transparency.
In contrast, a coexistent orthorhombic-tetragonal phase structure
at room temperature is responsible for its superior piezoelectric
performance, which outperforms other reported KNN-based transparent
ceramics \citep{Li2023}. Krishnaswamy et al. \citep{Krishnaswamy2019a}
also concentrated on increasing the performance of lead-free piezoelectric
composites by using poly-crystalline inclusions and modifying the
dielectric matrix environment. In such circumstances, their study
demonstrates that polycrystalline piezoelectric materials can outperform
their single crystal counterparts in isolation, but these improvements
could be more obvious in a composite design. They identify the causes
of loss that prohibit polycrystalline inclusions from improving composite
performance. They demonstrate an enhancement of 14.6\% in piezoelectric
response by polycrystalline inclusions. Lead-free piezoelectric materials
with higher thermal stability and enhanced electromechanical properties
are desired. Therefore, lead-free material 0.915(K$_{0.45}$Na$_{0.5}$Li$_{0.05}$)NbO$_{3}$-0.075BaZrO$_{3}$-0.01(Bi$_{0.5}$Na$_{0.5})$
modified with Sm$^{3+}$ ceramics made of TiO$_{3}$ (KNLN-BZ-BNT)
are created. The optimal piezoelectric constant values of 324 and
385 pC/N are obtained with an enhanced temperature range (30-180$\lyxmathsym{\textcelsius}$)
\citep{Quan2021}. Using finite element analysis, PZT, KNLNTS, and
BNKLBT are studied for active vibration control. It is found that
KNLNTS have better performance than lead-based materials for actuators.
This may be because KNLNTS is stiffer than PZT material, resulting
in faster damping and vibration control \citep{Sharma2014}.

Moreover, the development of next-generation sophisticated piezoelectric
transducers is heavily dependent on the use of lead-free piezoelectric
ceramics. In this study, K$_{0.5}$Na$_{0.51}-x$Ca$_{x}$NbO$_{3}$ ceramics,
denoted as $(100-x)$ KNCN, are synthesized using a solid-state reaction
technique. The influence of defect dipoles and oxygen vacancies on
the electromechanical properties of potassium sodium niobate (KNN)
ceramics is investigated. The findings of this study reveal that the
addition of a suitable amount of calcium ions can establish a border
between the rhombohedral and tetragonal phases (referred to as the
R-O phase boundary), leading to enhancements in the d$_{33}$ and
Q$_{m}$ properties. In addition, the achievement of a desirable equilibrium
between electromechanical properties and piezoelectric coefficient
is attained through the process of annealing the sample in a nitrogen
(N$_{2}$) atmosphere, specifically using a combination of 2KNCN
and N$_{2}$. In comparison to 2KNCN, the d$_{33}$ and Q$_{m}$
values of 2KNCN/N$_{2}$ exhibit an approximate increase of 30\%
and 41\%, respectively. The objective of this study is to propose
a robust framework for defect engineering with the aim of establishing
its viability as a viable option for industrial applications in the
field of piezoelectric materials \citep{Qi2023}.

Ferroelectric ceramic-based materials demonstrate piezoelectricity,
electro-optics, acousto-optics, and nonlinear optical properties.
The lead-free ceramic was synthesized by the process of mechanical
grinding. The tetragonal and orthorhombic phases shown by a lead-free
ferroelectric ceramic material with the chemical composition K$_{0.35}$Na$_{0.65}$Nb$_{0.97}$Sb$_{0.03}$O$_{3}$.
In addition to conducting structural studies by Rietveld analysis, the
chemical composition of the items was determined using VEELS and CEELS
techniques. The peak corresponding to the distinctive characteristics
of perovskite structures becomes evident when the temperature reaches
950${^\circ}C$ \citep{LealPerez2023}. The high temperature tetragonal
phase distributions limit the formation of orthorhombic phases. It
is demonstrated, in particular, that two main mechanisms regulate
the reduction of stress processes in which structural and microstructural
effects are correlated; the first is associated with a purely microstructural
development in which bimodal grain distribution inhibits the formation
of non-$180{^\circ}$ domains. Conversely, the second is primarily
regulated by ferroelectric domain distribution and the presence of
pseudo-cubic areas around room temperature linked with local structural
heterogeneity (polar nano-regions, PNRs). These processes, in particular,
produce areas with various stress states and explain the expansion
of the phase transition owing to phase coexistence \citep{Moure2023}.
Further, MnO$_{2}$-doping can increase the development and electrical
characteristics of K$_{0.5}$Na$_{0.5}$NbO$_{3}$-BiAlO$_{3}$ (KNN-BA) single
crystals. Furthermore, at $x=0.005$, the size of the $(1-x)$(KNN-BA)-$x$MnO$_{2}$
crystals reached a maximum of around 18$\times$17$\times$2 mm$^{3}$; due
to the addition of MnO$_{2}$, the colour of the crystals progressively
changed from bright yellow to dark brown. The XRD and refinement data
revealed that adding MnO$_{2}$, in the range of $0<x<0.009$, did
not affect the crystalline structure of the perovskite crystals. All
crystals have a two-phase perovskite crystalline structure with orthorhombic
and tetragonal phases. With an increase in MnO$_{2}$, the orthorhombic
phase content dropped while the tetragonal phase content rose. Furthermore,
the domains in the crystals were mainly $180{^\circ}$ and had a parallel
lamellar structure. In addition, crystals with $x=0.003$ had the
shortest average domain width. These crystals exhibited the highest
piezoelectric constant (568 pC/N) and overall performance \citep{Wang2023}.

\subsection{Bismuth sodium titanate (BNT)-based piezoelectrics}

The BNT and BNT-based piezoelectric ceramics demonstrate robust ferroelectric
properties, including a significant remnant polarization of P$_{r}=$38$\mu$ C/cm$^{2}$,
and superior piezoelectric attributes when compared to lead-free piezoelectric
ceramics. As a result, BNT is seen as a good contender for use as
a crucial material in lead-free piezoelectric ceramics \citep{chen2024,cai2024}. However, data
on the piezoelectric characteristics of the BNT ceramic are sparse
since, except in specialized work, it is impossible to pole this ceramic
due to a strong coercive field (E$_{c}=$73kV/cm) \citep{Nagata2006}.
Further, in a site-based study, it is found that BNT can be poled
easily, and the electromechanical coupling factor, k$_{33}$, of the
BNT ceramic as the end-member of solid solutions ranges from 0.25
to 0.40 because of differences in sintering circumstances. The BNT
ceramic requires a high sintering temperature of more than 1200 ${^\circ}C$
to get a dense body. Bi-ion vaporization is predicted to occur during
the sintering process at temperatures more than 1200 ${^\circ}C$,
leading to poor poling treatment due to low resistivity. According
to the thermograph (T$_{G}$, weight loss) measurement, the loss due
to Bi vaporization occurred at temperatures over 1130 ${^\circ}C$.
Many techniques and methods are proposed to prevent Bi vaporization
and achieve stoichiometric BNT ceramic. As a result, the BNT ceramic
should be sintered at 1100 ${^\circ}C$ or below \citep{Herabut1997,Devi2016}.
For improved piezoelectric and dielectric characteristics, $(1-x)$BNT$-x$KNN$(x=0-0.02)$
and 1 wt\% Gd$_{2}$O$_{3}$ are studied. The inclusion of KNNG composition
resulted in a decrease in the T$_{d}$ and T$_{C}$ values. The frequency
dispersion in T$_{C}$ demonstrates the relaxor behaviour of BNT-KNNG
ceramics. The leakage current density was significantly lowered by
replacing KNNG in the BNT system, and the dielectric and piezoelectric
characteristics were enhanced. According to the results, high-power
electromechanical applications may benefit from BNT-KNNG $(x=0.01)$
ceramics \citep{Pattipaka2018}. The high-temperature performance
of lead-free BNT-type piezoelectric material is enhanced by shifting
the phase structure from R to RT coexisting phase; a steady operating
temperature of 150${^\circ}C$ was attained. This is accomplished
by incorporating PMN and BST into a non-stoichiometric BNT matrix,
which increases the material's ferroelectric relaxor by reducing the
residual polarization while maintaining the maximum polarization \citep{Yang2023}.

It has been demonstrated that in quenched BNT-7BT ceramics, the depolarization
temperature may be raised by 44 ${^\circ}C$ while maintaining a temperature-independent
d$_{33}$ throughout a wide temperature range of 25-170 ${^\circ}C$. We
concluded that the built-in field produced by simple quenching encourages
the improvement of the ferroelectric state connected to the R3C phase,
which is critical for increasing the T$_{d}$ and preserving the thermal
stability of d$_{33}$. Therefore, the synthesis of E$_{b}$ in quenched
BNT- 7BT ceramics has offered a new approach to addressing the conflict
between temperature stability, T$_{d}$, and d$_{33}$ in lead-free
piezoceramics \citep{Chen2022}. Furthermore, 1 mol\% AlN elevates
d$_{33}$ from 165 to 234 pC/N and raises T$_{d}$ by 50 ${^\circ}C$.
Rietveld analysis of X-ray powder diffraction (XRD) data showed an
increase in the quantity of the tetragonal phase at 1 mol\% AlN inclusion.
Furthermore, the modified ceramics feature larger grains and high-density
lamellar nano-domains with widths ranging from 30 to 50 nm with
this composition. As a result, polarization reversal and domain mobility
are considerably increased, resulting in the massive d$_{33}$. Temperature-dependent
dielectric and XRD analyses revealed that the delayed thermal depolarization
in the modified ceramics is caused by an improved and poling-field
stable tetragonal structure \citep{Zhou2021}. The piezo-catalytic
activity of Ag/BNT-AN piezo-catalyst was considerably increased by
combining the enhancement of piezoelectricity by Ag and Nb co-substitution
with the promotion of O$_{2}$ reduction and H$_{2}$O oxidation processes
by surface modification with Ag nanoparticles and hydroxyl groups.
Under ultrasonic vibration, the ideal Ag/BNT-AN piezo-catalyst produced
469 $\mu$ mol/gh, 13 times greater than pure BNT. Furthermore, when
the catalyst was subjected to simple stirring as mechanical stress,
H$_{2}$O$_{2}$ could be synthesized, showing the potential of Ag/BNT-AN
to capture low-frequency mechanical energy. This study proposes a
kilogram-level catalyst design paradigm and an effective technique
for controlling piezo-catalysts \citep{Zhang2023}.

Among other experimental techniques used for piezoelectrc materials, we should mention the electrospinning method \citep{Ji2016}. For example, by electrospinning flexible lead-free piezoelectric nanofiber
composites of BNT-ST (0.78Bi$_{0.5}$Na$_{0.5}$ TiO$_{3}$-0.22SrTiO$_{3}$)
ceramic and PVDF polymer. The crystal structure, microstructure, and
morphology of the as-prepared BNT-ST/PVDF nanofibers composite with
varied BNT-ST concentrations ranging from 0\% to 80\% were investigated
using XRD, FE-SEM, and EDS. The polarization-electric field (P-E)
loops were used to analyze the piezoelectric characteristics, and
the findings showed that a BNT-ST content of 60\% had improved piezoelectric
elements. The potential of frequency sensing application was proven
by evaluating the output voltage as a function of frequency of the
BNT-ST/PVDF nanofiber composite with a BNT-ST content of 60\% \citep{Ji2016}.
Hard-type piezoelectric ceramics that are environmentally sustainable
are in great demand for high-power transducer applications, replacing
lead-based piezoelectric ceramics. Piezoelectric material without
lead $(1-x)$(0.8Bi$_{0.5}$Na$_{0.5}$TiO$_{3}$-0.2Bi$_{0.5}$K$_{0.5}$TiO$_{3}$)
On Pt$_{(111)}$/Ti/SiO$_{2}$/Si substrates, $(1-x)$Bi(Ni$_{0.5}$Zr$_{0.5}$)O$_{3}$
thin films (abbreviated as BNT$-$BKT$-x$BNZ) $(x=0.00,0.01,0.02,0.03,0.04)$
were made using the sol-gel process. A thorough investigation was
also conducted into the effects of Bi(Ni0.5Zr0.5)O3 content on the
microstructure, dielectric, ferroelectric, and piezoelectric characteristics.
It was discovered that the Bi(Ni$_{0.5}$Zr$_{0.5}$)O$_{3}$ composition
significantly impacted the rise in relaxor and the fall in oxygen
vacancies, which are essential for enhancing thin-film characteristics.
With polarisation of 40.27C/cm$^{2}$, dielectric constants of 477,
and an effective inverse piezoelectric coefficient of up to 125.9
pm/V, the thin film of BNT-BKT-0.02BNZ demonstrated the best electrical
characteristics. The BNT-BKT thin films with 0.02 mol\% Bi(Ni$_{0.5}$Zr$_{0.5}$)O$_{3}$-doped
were found to be a type of lead-free piezoelectric material with outstanding
manifestations and a promising future for application development
\citep{Ni2022}. 

Moreover, Bi$_{0.5}$Na$_{0.5}$TiO$_{3}$ piezoelectrics must be thermally stable for practical applications. Adding Ag to the perovskite 0.94Bi$_{0.5}$Na$_{0.5}$TiO$_{3}$-0.06BaTiO$_{3}$ The (BNT-6BT) matrix experiences an increase in tensile stress after cooling from sintering temperatures due to a large thermal expansion coefficient difference. The FEM study led to this finding. Silver (Ag) is added to BNT-6BT and composite ceramics with a perovskite/metal structure (BNT-6BT/100$x$Ag) to raise the depolarization temperature ($T_{d}$). The mix with $x = 0.06$ has a thermal decomposition temperature ($T_{d}$) of 145 $^{\circ}C$, which is 45 $^{\circ}C$ higher than the BNT-6BT prototype and doesn't change its piezoelectric response. Research shows that residual thermal stress, particle size beyond 200 $\mu m$, and oxygen vacancies all raise $T_{d}$ \citep{Luo2023a}. Furthermore, high performance, compactness, and simple integration with semiconductor processes have made piezoelectric micromachined ultrasonic transducers (PMUTs) popular. A simulated Bi$_{0.5}$Na$_{0.5}$TiO$_{3}$ piezoelectric single-crystal film was used to make the PMUT. It has a high frequency, a larger electromechanical coupling factor ($k_{eff}^2$), and a high sensitivity. As a result, it is well-suited for medical ultrasound imaging that requires high resolution and high frequency. The BNT-BT piezoelectric single-crystal film's static sensitivity and $k_{eff}^2$ are both improved by changing the Euler angle $\gamma$. This gives it a value of 1.34 between 45$\pm$5$^{\circ}$. The spinning BNT-BT in the PMUT gave it a relative pulse-echo sensitivity of -46 $dB$ and a bandwidth of 35\% at a 200 μm reflecting surface \cite{Liu2023}.

\subsection{Barium titanate (BT) based piezoelectrics}

Lead-free piezo composites with CNT-modified matrix have been studied
by Krishnaswamy et al. \citep{A.Krishnaswamy2019}, and they found
2-3 orders of magnitude increase in piezoelectric response by simultaneously
hardening the matrix and improving its permittivity by adding CNT
in the matrix. They discovered that near nanotube percolation, nanotube
clustering can lead to more significant matrix hardening and higher
permittivities, resulting in more than 30\% piezoelectric response
increases compared to non-agglomerated structures. They claimed that
altering the polycrystalinity of piezoelectric inclusions improved
piezoelectric responsiveness by more than 50\%. It has also been discovered
that the atomic vacancy defect in nanotubes softens the matrix, which
may be enhanced by agglomerating nanotubes at higher length scales.
Further, Krishnaswamy et al. \citep{Krishnaswamy2019} added graphene
nanotubes in PVDF matrices with BT as piezoelectric inclusions and
found that the mechanical properties are superior to CNT inclusions.
However, the material hardening is half in the case of graphene nano-inclusions
compared to CNT inclusions. They go on to explore the influence of
the electric flux and fields in the graphene-modified piezo composite
on the polycrystalinity of the piezoelectric inclusions to discover
polycrystalline configurations that might lead to increased performance
in such nano-modified piezo composites.

Several CaHfO$_{3}$ modified BiFeO$_{3}$-0.33BaTiO$_{3}$(BF-0.33BT-$x$CH)
lead-free piezoceramics were produced using the solid-state sintering
technique. The elimination of core-shell structures featuring non-uniform
element distributions in their compositions can be achieved through
the addition of a significant amount of CH content, specifically when
$x$ exceeds 0.05. The BF-0.33BT-0.01CH piezoceramic exhibits the
highest saturation polarization (40.1$\mu$C/cm$^{2}$), remnant polarization
(26.8$\mu$C/cm$^{2}$), and converse piezoelectric coefficient (290
pm/V), which are notably improved compared to the undoped BF-0.33BT
piezoceramic. The piezoelectric capabilities of BF-0.33BT-$x$CH ceramics
demonstrate a quick fall with an increase in CH content, leading to
the manifestation of relaxor ferroelectric features \citep{Zhao2023}.
Moreover, due to its high Curie temperature (T$_{C}$) and superior
piezoelectric performance, lead-free BiFeO$_{3}$-BaTiO$_{3}$ ceramics
have received much interest in the past 20 years. Here, using a cutting-edge
poling technique (AC-bias $+$ DC-bias) with a high T$_{C}$ of 455 ${^\circ}C$,
it was possible to produce an excellent piezoelectric constant (d$_{33}$)
of 325 pC/N in the Nd-modified 0.67BiFeO$_{3}$-0.33BaTiO$_{3}$ ceramics.
Additionally, in the temperature range of 25 to 125 ${^\circ}C$,
a very high normalized piezoelectric strain (d$_{33}$=S$_{max}$/E$_{max}$;
where S$_{max}$ is the maximum mechanical strain, and E$_{max}$
is the maximum electric potential) of 808 $pm/V$ was produced at
the average/typical and relaxor-ferroelectrics phase border concurrently
with thermal solid stability ($\Delta$ d$_{33}$(T)$=$20\%) \citep{Habib2023}.
Further, piezoelectric ceramics made of 0.67Bi$_{1.03}$FeO$_{3}$-0.33BaTi$_{1-x}$Zr$_{x}$O$_{3}$
(BF0.33BT-100$x$Zr with $x=0.00-0.05$) were examined for their structural
evolution and electrical property change. Near the morphotropic phase
boundary (MPB), it was possible to get an improved static piezoelectric
coefficient (d$_{33}$) of 248 pC/N with a high Curie temperature
(T$_{C}$) of roughly 400 ${^\circ}C$. As the temperature was raised
to 100${^\circ}C$, it was discovered that the dynamic piezoelectric
coefficient (d$_{33}^{*}$) at room temperature steadily rose from
460 to 620 pm/V. The BF0.33BT-3Zr ceramic demonstrated a significant
electrostrictive coefficient (Q$_{33}$) in the 25\textendash 100
${^\circ}C$ temperature range of 0.018\textendash 0.032 m$^{4}$/C$^{2}$.
The thermal facilitation of domain switching is related to the amplification
of the ferroelectric and piezoelectric responses of the ceramic with
rising temperature \citep{Habib2022}. MnO doping decreased the rhombohedral
phase's lattice distortion and aided the pseudo-cubic phase's emergence
in the 0.75BiFeO$_{3}$-0.25BaTiO$_{3}$ system. Furthermore, it resulted
in domain morphological variety and domain size reduction. XPS study
revealed that Mn$^{2+}$ ions inhibited the transition from Fe$^{3+}$
to Fe$^{2+}$ ions, increasing oxygen vacancies. As a result, adding
MnO raises the d$_{33}$, K$_{p}$, and Q$_{m}$ while decreasing
the $tan \delta$ of 0.75BiFeO$_{3}$-0.25BaTiO$_{3}$ ceramics. The aberrant rise
in hard-type characteristics is mainly attributed to an increase in
extrinsic contribution rather than a pinning effect caused by oxygen
vacancies \citep{Luo2023}. 

\subsection{Li-doped lead-free piezoelectric materials}

Certain studies have shown lithium doping to raise the Curie temperature
(T$_{C}$) and decrease the phase transition orthorhombic-tetragonal
temperature (T$_{o}$-T) to a lower value, close to room temperature
{[}19{]}. While improving the performance of the piezoelectric characteristics,
low-temperature melting additives like lithium (Li$_{2}$CO$_{3}$, LiF)
can also aid in lowering the sintering temperature \citep{Fadhlina2022}.
However, since Li$_{2}O$ has a higher melting point (over 1700$\lyxmathsym{\textcelsius}$),
the doping process must be altered to produce a greater density and
impact the material's crystallinity \citep{Ma2012}. Fully crystalline
materials have a greater density than other materials in the secondary
phase. Aside from that, lithium, Li doping on piezoelectric materials,
can alter the structure. A particular quantity of lithium doped on
KNN-based material results in the coexistence of orthorhombic-tetragonal
phase structures at room temperature \citep{Chen2016}. As a result,
the piezoelectric characteristics of that material are enhanced. Material
with orthorhombic-tetragonal phase coexistence offers more excellent
piezoelectric characteristics than material with solely orthorhombic
or tetragonal phases \citep{Zhang2015}. For all electrical parts,
the material 0.995(K$_{0.5}$Na$_{0.5}$)1-xLi$_{x}$NbO$_{3}$-0.005BiAlO$_{3}$
with $x=$0.00,0.01,0.02,0.03,0.04,0.05,0.06, and 0.08 yields an inconsistent
result. However, the electrical property coefficients have the most
significant value when the material is at the orthorhombic-tetragonal
phase border \citep{Chen2017}. On the other hand, the coexistence
of the rhombohedral-tetragonal phase boundary was shown to be the
most beneficial in increasing the piezoelectric response \citep{Li2020}.

However, deformation of phase structures can occur with specific quantities
of lithium doping, as seen by the band gap. The band gap decreases
as the concentration of Li doping rises in 0.97Bi$_{0.5}$Na$_{0.4}-x$Li$_{x}$K$_{0.1}$TiO$_{3}$-0.03CaZrO$_{3}$
(BNKTCZ-$x$Li) with $x=$0.00,0.01,0.02,0.03,0.04, and 0.05. According
to Quyet et al. \citep{vanQuyet2015}, one of the causes of the lower
optical band gap is a deformed structure. To improve piezoelectric
characteristics, lithium doping can move the polymorphic phase transition
closer to ambient temperature \citep{Tan2013}. Aside from that, lithium
dopants have been frequently utilized since they are less expensive
than other dopants. The doping of lithium in lead-free piezoelectric
BCTZ. has enhanced its electromechanical properties tremendously.
A doping of 0.5 wt \% of Li$_{2}$CO$_{3}$ has increased the value of
the piezoelectric coefficient (d$_{33}$) up to 530 pC/N, which
is a huge increment \citep{Fadhlina2022}. Investigations were done
into how low lithium substitution affected the electrical characteristics
and microstructure of (K$_{0.5}$Na$_{0.5}$)$_{(1-x)}$Li$_{x}$NbO$_{3}$ (KNLN)
ceramics. The sol-gel technique was used to synthesize each sample.
The monoclinic-tetragonal (T$_{M}$-T) phase transition gradually
slowed as the ceramics' Curie temperature (T$_{C}$) increased. Li$^{+}$
substitution significantly increased the piezoelectric coefficient
(d$_{33}$) up to 181 pC/N and greatly impacted the ferroelectric
characteristics \citep{JanilJamil2022}.
\begin{figure} [t]
    \centering
    \includegraphics[width=0.8\linewidth]{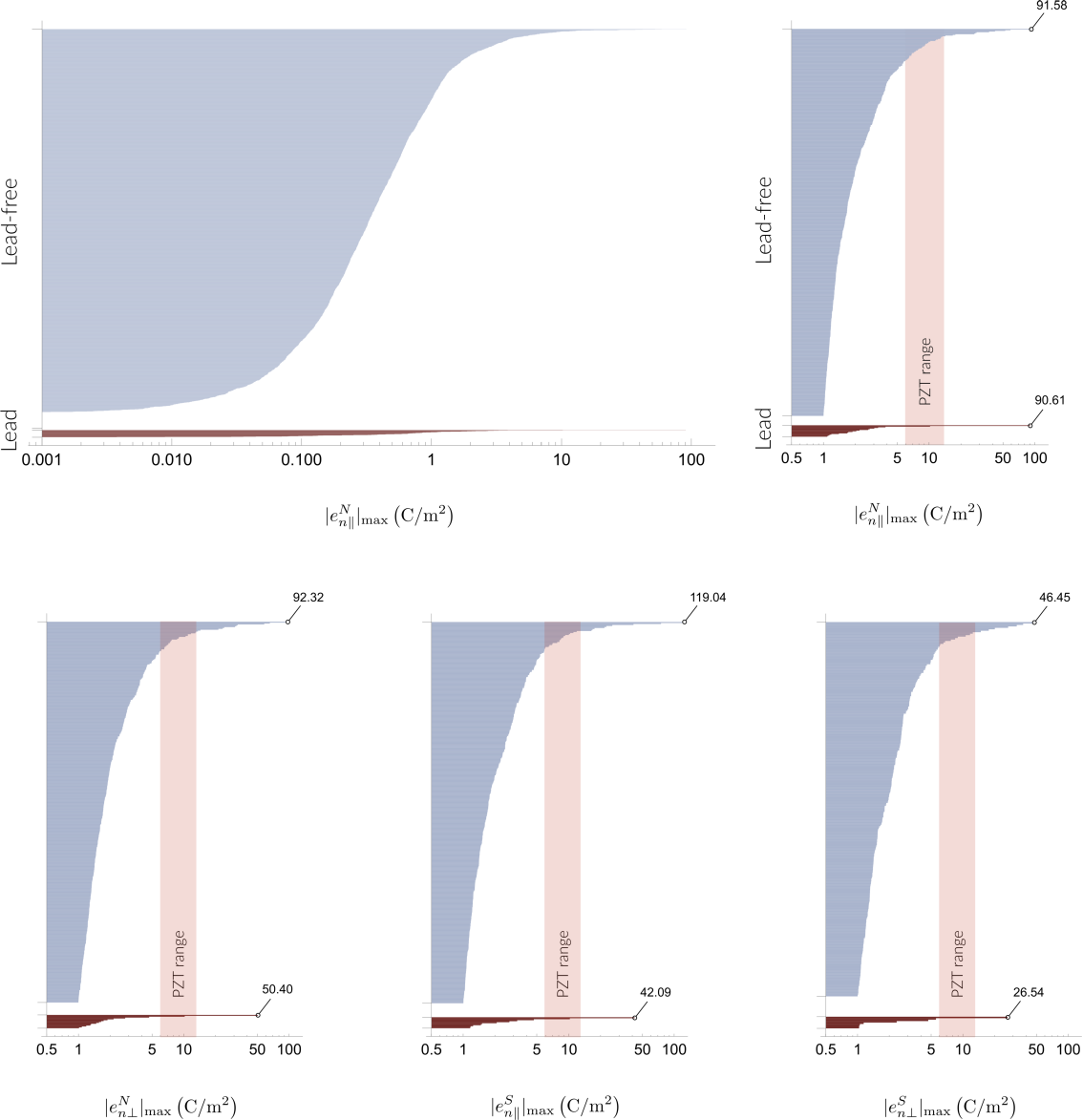}
    \caption{Bar charts showing the four Figures of Merit for the piezoelectric coefficients of all materials analyzed. The top-left plot presents the maximum longitudinal piezoelectric coefficient, $|e^N_{n\parallel}|_{\max}$, for all materials, while the top-right plot zooms in on values exceeding $1~\mathrm{C}/\mathrm{m}^2$. The bottom row provides details for $|e^N_{n\perp}|_{\max}$, $|e^S_{n\parallel}|_{\max}$, and $|e^S_{n\perp}|_{\max}$.}
    \label{fig:FigE}
\end{figure}

\subsection{Advancing materials discovery through databases}

 A promising strategy to accelerate the discovery of lead-free piezoelectric materials is leveraging computational tools and atomistic simulations. The Materials Project (MP, \url{www.materialsproject.org}), aligned with the Materials Genome Initiative, is a comprehensive platform designed to use high-throughput first-principles calculations for predicting a wide range of material properties~\cite{MP}. This initiative has become a cornerstone in materials design and discovery. Within this broader framework, the work of de Jong et al.~\cite{deJong2015}  focuses specifically on piezoelectric constants, significantly expanding the database of computed piezoelectric tensors. By identifying key structural and compositional factors influencing piezoelectric performance, their contribution exemplifies the Materials Project’s capacity to inform the development of predictive models. Such models provide invaluable insights and tools for designing next-generation lead-free piezoelectric compounds with enhanced functionality and sustainability.

The piezoelectric properties in the MP database are obtained using first-principles quantum-mechanical calculations based on Density Functional Theory (DFT), particularly Density Functional Perturbation Theory (DFPT)~\cite{deJong2015}. These calculations yield intrinsic piezoelectric constants associated with bulk, defect-free, and strain-free materials at a temperature of 0~K. However, the database presented by de Jong et al.~\cite{deJong2015} does not differentiate lead content, and since the paper’s publication, many new materials have been incorporated. Originally, their study featured 941 materials, but the database has since expanded under the Materials Project framework. In this study, we extend the MP database by incorporating additional properties via post-processing DFT data. Our screening includes a total of 3,292 piezoelectric materials. During this process, the database was curated to identify lead (Pb) content and to categorize materials as either lead-based (58 materials) or lead-free (3,234 materials). The database was accessed on August 21, 2024, via the MP API.

From the piezoelectric properties provided by the MP API, we computed four Figures of Merit based on the piezoelectric stress coefficients. The longitudinal piezoelectric coefficient, denoted as $e^N_{n\parallel}$, measures the electric displacement in a direction $\bm{n}$ where the strain is applied parallel to this direction. The subscript $\parallel$ indicates that the strain is aligned with $\bm{n}$, while the superscript $N$ refers to a normal strain. Conversely, the subscript $\perp$ indicates that the normal strain is applied perpendicular to the direction $\bm{n}$. When the superscript is $S$, the strains considered are shear strains, with $\parallel$ and $\perp$ referring to out-of-plane and in-plane strains, respectively.

Figure \ref{fig:FigE} presents the four Figures of Merit for each compound as bar charts. Note that this is not a cumulative distribution; the data are simply sorted for clarity. At the top, the maximum longitudinal piezoelectric coefficient, $|e_{n \|}^N|_{\max}$, is shown for all materials. The top-right subfigure focuses on data above $1~\mathrm{C}/\mathrm{m}^2$, highlighting the maximum values for lead-free and lead-based materials, which are $91.58~\mathrm{C}/\mathrm{m}^2$ and $90.61~\mathrm{C}/\mathrm{m}^2$, respectively. Notably, more than $97\%$ of materials exceeding this threshold are lead-free. For comparison, lead zirconate titanates (PZTs) exhibit maximum absolute piezoelectric tensor components in the range of approximately $6-12~\mathrm{C}/\mathrm{m}^2$~\cite{deJong2015}. The bottom row of Figure \ref{fig:FigE} provides detailed data for the maximum $|e^N_{n\perp}|$, $|e^S_{n\parallel}|$, and $|e^S_{n\perp}|$.

While many materials in the MP database remain unsynthesized, these results highlight the untapped potential of lead-free compounds as sustainable, high-performance alternatives to lead-based piezoelectrics, with promising opportunities for application-specific optimization.

This section has presented an overview of specific steps for improving the performance of
the most prominent lead-free piezoelectric materials through such innovative techniques as phase boundary engineering (PBE) and other creative multiscale designs of piezoelectric composites. The improved performance by these innovative techniques of environmentally friendly piezoelectric materials has a significant impact on a wide range of applications. The following sections
present the information on most important enhancement techniques used to different
piezoelectric materials and their impact of improved performance
for certain applications.

\section{Applications of lead-free piezoelectric materials}
\markright{Applications of lead-free piezoelectric materials \hfill \empty}

Piezoelectric materials have been employed in a variety of applications.
Some prominent examples include the nanogenerator application by Barium Titanate
(BT) and Barium Calcium (BC) material was owing to high power density
(0.64 $\mu$ W/cm$^{2}$) \citep{Wei2018}; however, a hydrothermal approach
was used to apply ZnO at polyvinylidene fluoride (PVDF) material for
wearable nanogenerator use \citep{Kim2018}. Aside from nanogenerators,
lead-free piezoelectric materials are employed in medical tools as
transducers. The end product of (K,Na)NbO$_{3}$-KTiNbO$_{5}$-BaZrO$_{3}$-Fe$_{2}$O$_{3}$-MgO
(KNN-NTK-FM) is a biomedical ultrasonic imaging material with high
frequencies (52.6 MHz), a broad bandwidth (64.4\%), and an excellent
electromechanical coupling coefficient (k$_{p}\sim$ 0.45) \citep{Chen2019}. 

Because of their environmental friendliness and potential for a wide
range of applications, lead-free piezoelectric materials have received
much interest. They have comparable piezoelectric capabilities to
standard lead-based materials but without the harmful consequences
of lead. The following are some of the most essential uses for lead-free
piezoelectric materials.

\begin{figure}
    \centering
    \includegraphics[width=1.0\linewidth]{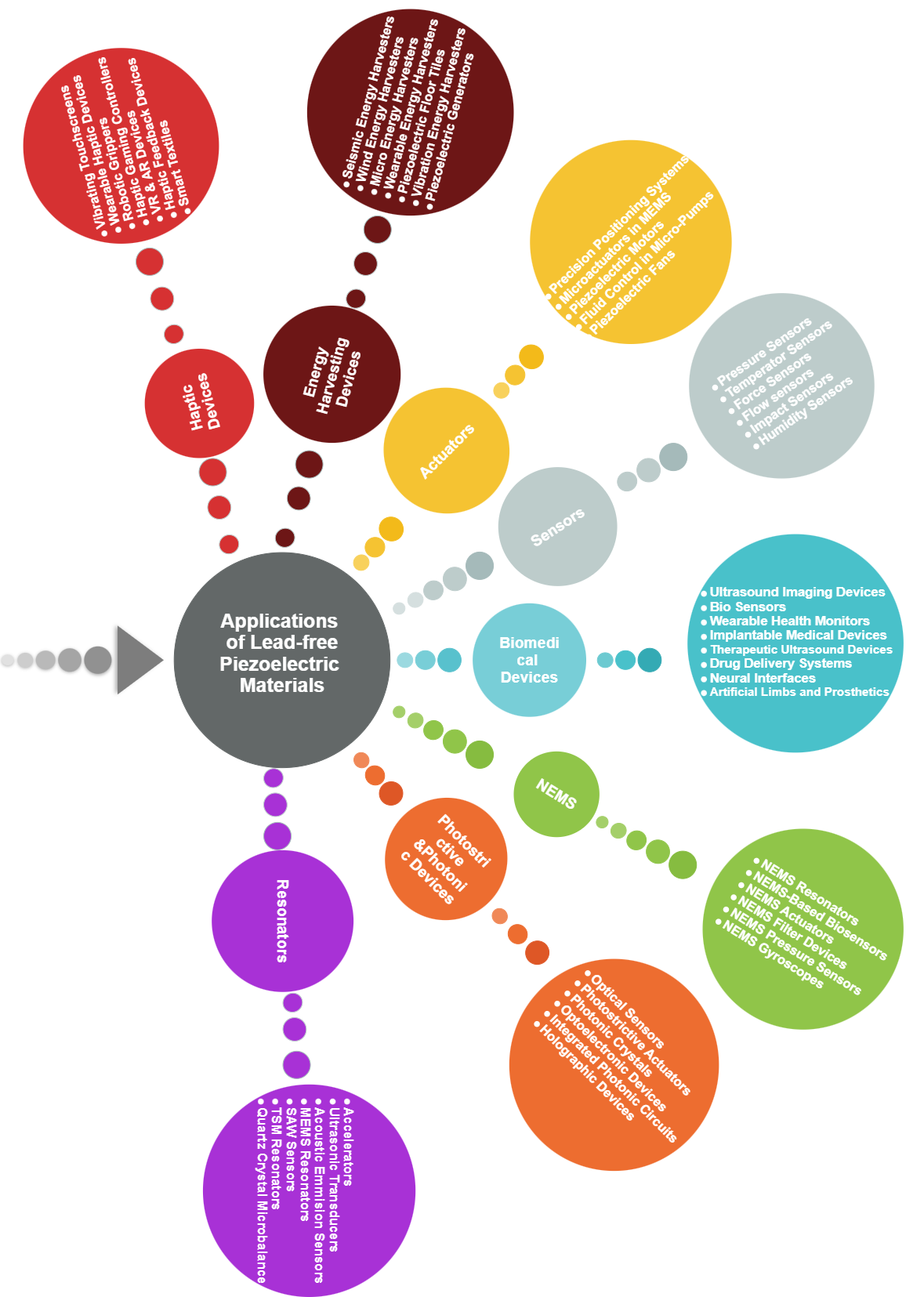}
    \caption{Schematic diagram of various applications of lead-free piezoelectric materials}
    \label{fig:Schematic2}
\end{figure}

\subsection{Actuator applications}

Actuators often need piezoceramics with a high Curie temperature (T$_{C}$),
a sizeable electric-field-induced strain, and a significant blocking
force (force generated in the wholly clamped condition). These settings
primarily define the actuator's functioning range and are changed
to meet the needs of each unique device.

Murata Manufacturing Co., Ltd. of Japan was at the forefront of developing
lead-free multilayer piezoceramic actuators with base-metal inner
electrodes \citep{Kawada2009,Kawada2015}. They discovered that lead-free
piezoceramics based on KNN may be densely sintered at low oxygen partial
pressure. Using a unique formula, they constructed a 12-layer multilayer
piezoelectric actuator with the most significant recorded electric-field
induced normalized strain, Smax/Emax, at 2 kV/mm. Although this
value is only approximately half that of PZT-based actuators, it is
hypothesized that by adding more piezoceramic layers and decreasing
their thickness, KNN-based multilayer ceramic actuators might attain
the same displacement as PZT-based actuators. Further, Noliac A/S
of Denmark created lead-free multilayer actuators with relaxor ferroelectrics
\citep{CarrenoJimenez2018} utilizing BNT-based piezoceramics. While
their strain response was lower than that of PZT-based actuators,
they demonstrated outstanding cycle stability in numerous formulations.

The high values of piezoelectric response and strain responsiveness of BNT can be harnessed by increasing its density by doping it with other materials. Nguyen et al. \citep{quyet2015} created ceramics by altering bismuth sodium potassium titanate with lithium and calcium zirconate, culminating in the formation of Li-modified bismuth sodium potassium titanate-Calcium zirconate (BNKT-CZ) ceramics. When subjected to an electric field strength of 60 kV/cm and containing a lithium ion concentration of 4 mol\%, these ceramics displayed a strain of 0.40\%. The ceramics had strain characteristics that were twice as large as those of an undoped composition. Nguyen et al. \citep{nguyen2012} reported that the strain of bismuth sodium potassium tin titanate (BNKSnT) ceramics was 0.39\%. These ceramics were doped with 4 mol\% Li$^{+}$ and were positioned in close proximity to the phase transition point. Additionally, the researchers observed an improvement in the density of the ceramics that were produced by utilizing this combination, in addition to an improvement in the formation of the grain structure. For the purpose of producing Sn-modified BNKT, Xi et al. \citep{xi2020} employed the solid-state reaction technique as their method of experimentation. Ceramics exhibit a transition from a tetragonal structure to a pseudo-cubic relaxor type when the amount of Sn$^{4+}$ present in the material is increased by up to 0.02 mil. The composition was found to have a maximum strain of 0.37\%, and the P-E hysteresis loop was found to be quite narrow.

\subsection{Sensor applications}

Piezoelectric sensors convert external mechanical strain into electrical
signals, allowing the detection of various acoustical impulses or
pressures \cite{sadl2018}.

NGK Spark Plug Co., Ltd. of Japan created a spray-drying composite
system combining KNN-based piezoceramics and KTiNbO$_{5}$ (KNN-KTN)
\citep{Matsuoka2014}. The material is offered in 100-kg increments
and has a high planar coupling coefficient (k$_{p}$) of 0.52 while
age-resistant. This mixture was used to make a knocking sensor for
automobile engines, which is in charge of monitoring the combustion
process and modifying ignition timing. The lead-free sensor has the
same output voltage as the PZT sensor and is now available on the
market. Output voltage and endurance superior to those currently available
on the market produced using PZT. Another important application is
the self-power temperature sensor based on flexible thermoelectric generator (f-TEG), which may immediately transmit
an electrical signal without any transduction, allowing for easy integration
of these temperature sensors with wearable electronics for real-time
health monitoring \citep{Yang2019,Zhao2022}.

Furthermore, multifunctional sensor systems may be realized when f-TEG
is combined with other sensing units with various functionalities
\citep{Zhang2015a,Wang2020}. We recently proposed a strategy for
creating a vertically configured flexible pressure/temperature bimodal
sensor array using a 3D architecture design on all-organic thermoelectric and
piezoelectric materials, demonstrating fast, highly sensitive, high
selectivity, and wide sensing range tactile sensing ability in a practical
human-machine interaction scenario. The bimodal tactile sensor array enables the precise mapping of pressure and temperature signals concurrently, without any interference. Furthermore, this sensor operates autonomously without the need for an external power source, according to its inherent operating principle \citep{Zhu2020}. The findings promote the progress of flexible thermoelectric generators (f-TEGs) for wider use in multifunctional electronic skin (e-skin) \citep{Wang2023a}.

Furthermore, an electrospinning technique was used to create flexible composites of BNT-ST (0.78Bi$_{0.5}$Na$_{0.5}$
TiO$_{3}$-0.22SrTiO$_{3}$) ceramic and polyvinylidenefluoride (PVDF) polymer, which are lead-free and piezoelectric nanofibers. The BNT-ST/PVDF nanofibers composite, generated in various concentrations ranging from 0 wt\% to 80 wt\%, was analyzed using XRD, FE-SEM, and EDS to examine its crystal structure, microstructure, and morphology. The piezoelectric properties were examined using polarization-electric field (P-E) loops, and the findings indicated that a 60 wt\% BNT-ST composition exhibited improved piezoelectric characteristics. The feasibility of utilizing a frequency sensor was validated by measuring the output voltage in relation to the frequency of the BNT-ST/PVDF nanofiber composite containing 60 wt\% of BNT-ST \citep{JI2016316}.

\subsection{Biomedical applications}

In wearable applications, self-powered biocompatible electronic devices
that monitor critical physiological signals of the human body are
in high demand. A low-cost polydimethylsiloxane/polypyrrole (PDMS:PPy) composite
polymer film-based piezoelectric nanogenerator is built and used as
a self-powered arterial pulse pressure sensor. When
1.47 N/cm$^{2}$ pressure is applied, a flexible and resilient piezoelectric
nanogenerator with device configuration of Al-coated (PDMS:PPy) piezoelectric transducer (PET) produces an output voltage of 12 V and current density
of 0.11 $\mu$A/cm$^{2}$. This biocompatible nanogenerator is demonstrated
as a self-powered pulse sensor, demonstrating its potential for constructing
diverse self-powered systems for biomedical applications \citep{Veeralingam2022}.

In medical imaging, it is necessary for an ultrasonic transducer to possess a broad frequency range while still keeping adequate sensitivity.The initial prerequisite necessitates a brief duration of the pulse, leading to enhanced axial resolution. In order to strike a balance between competing factors, a transducer has to have a high electromechanical coupling coefficient (k$_{t}$), a low acoustic impedance (as similar as possible to that of biological tissue for optimal acoustic matching), and a suitable dielectric constant (for optimal electrical matching) \citep{Lethiecq2008}.
Furthermore, PVDF is utilized in medical imaging applications because to its low acoustic impedance. However, its limited electromechanical coupling factor restricts its usefulness in this field. Nevertheless, the availability of thin sizes enables high frequency (about 40 MHz) and high resolution imaging for the skin and anterior segment of the eye \citep{turnbull1995,silverman1995}. The poor electromechanical coefficient of PVDF can be overcome by utilizing its copolymers. In recent times, copolymers of PVDF have gained significant popularity as flexible piezoelectric elements in many advanced applications such as artificial muscles, soft robotics, haptics for virtual reality, and wearable sensors\citep{chen2022R}.

Moreover, other lead-free materials such as KNN, BNT, BT-based ceramics also exhibit significant values of elctromechanical coupling coefficient, low acoustic impedance and a suitable dielectric coefficients, which make them suitable for high and low frequency medical sensing applications \citep{Ringgard2024}. The BNT-BT 1\textendash 3 composite,
a piezoelectric composite, exhibits a significant electromechanical
coupling factor (k$_{t}=$0.71), which is comparable to that of the
lead-based Pb(Mg$_{\frac{1}{3}}$Nb$_{\frac{2}{3}}$)O$_{3}$-PbTiO$_{3}$/epoxy
1-3 piezoelectric composites at high resonative (31.88 MHZ) and antiresonative (43.16 MHz) respectively. This superior performance and lead-free
benefits of the BNT-BT/epoxy 1-3 piezoelectric composites make them
very appealing in high-frequency medical ultrasound applications,
particularly within the human body, such as intravascular ultrasound
imaging (IVUS) \citep{Zhou2022}.

Furthermore, synapse-based artificial neural networks (ANNs) provide potential solutions to the von Neumann bottleneck by enabling simultaneous processing and storage of input. They present an artificial synaptic device employing a ferroelectric BaTiO$_3$ thin film, distinguished by its robust weight update capabilities and diverse plasticity for neuromorphic applications. The effects of potentiation and depression are substantially affected by spike polarity, amplitude, frequency, amount, and rate. Furthermore, four variants of spike timing-dependent plasticities (STDP) and two types of Bienenstock-Cooper-Munro (BCM) learning rules with dynamic frequency thresholds have been established. In BCM learning rules, a conventional configuration demonstrates potentiation at elevated frequencies and depression at diminished frequencies under positive bias, whereas an unconventional design reveals depression at elevated frequencies and potentiation at diminished frequencies under negative bias. Furthermore, an ANN attains a recognition accuracy of 92.18\%. These results are essential for the potential application of ferroelectric artificial synapses in neuromorphic applications \citep{Yi20241}.

Moreover, the sensor may be utilized in electronic skin to enable precise and instantaneous collecting of temperature data. The sensing material consists of polyvinylidene fluoride-trifluoroethylene (PVDF-TrFE) and barium titanate nanoparticles \citep{gupta2017}. Thermistor sensors that are often used have limits regarding the proximity of the sensor to the human body. In response to this difficulty, flexible polymer temperature sensors have been developed. These sensors may be produced on thin, flexible polymer substrates employing adaptable printing methods such as inkjet, screen printing, and gravure printing \citep{ARMANKUZUBASOGLU2020}.

\subsection{Resonator applications}

Lead-free lithium niobate (LiNbO$_{3}$) piezoelectric transducers
are being examined as an alternative to lead-based systems for vibrational
energy scavenging applications. We implement a thick single-crystal
film on a silicon substrate with optimized clamped capacitance for
better impedance-matching conditions using a global optimization approach,
taking into account the significantly lower dielectric constant of
LiNbO$_{3}$ crystal compared to conventional piezoceramics (for example,
PZT). A power density of 8.26 W.mm$^{-2}$ is produced for a cantilever
with a bending first resonant frequency of 1.14 kHz, making it comparable
to lead-based and lead-free piezoelectric harvesters while utilizing
a widely used material and well-established production process (therefore
lowering the cost, for example). The device's energy-harvesting abilities
allow a sensor node to be started (from zero energy beginning circumstances)
after just 9 s and to maintain the ability to communicate data every
2 s (each transmission event using around 420 J) while being continuously
excited \citep{Clementi2021}.

\subsection{Haptic applications}

The touchscreens found on mobiles, laptops, smart watches, and automobile display screens are capable of recreating different human senses and providing sensory information to create virtual and augmented realities (VR/AR). These technologies aim to create a realistic and deeply engaging experience. However, the main obstacle preventing the widespread use of VR/AR devices in everyday life and other applications is the inflexible and bulky design of current technology, which does not accommodate the natural movements and flexibility of the human body. Recent advances in soft materials and wearable electronics are ideally
suited to provide answers to this dilemma. Because of the continual
heat exchange with other thermal entities via physical contact, thermal
sensibility, one of the tactile senses linked with human skin, performs
the essential function. The human body has the ability to distinguish objects with different thermal conductivities, even when they are at the same temperature, in addition to its basic sense of temperature. The Thermoelectric Generator (TEG) can be utilized for thermal management using the Peltier effect. This effect enables the active control of the device's temperature, allowing it to switch between cooling and heating by adjusting the magnitude and direction of the electrical current. This demonstrates the TEG's ability to regulate the desired temperature in a thermo-haptic system.

Lee et al. \citep{Lee2020} created a skin-like, highly soft and stretchable,
and bi-functional (both cold and hot sensation) thermo-haptic device
for wearable VR applications out of thermally conductive elastomer
encapsulated \textquoteleft p\textquoteright{} and n-type bismuth
telluride thermoelectric pellets connected by serpentine Cu electrodes.
They effectively demonstrated the collection of temperature data in a virtualized setting, where the user interacted with objects of different temperatures, from a cold beer bottle to a hot cup of coffee and from a warm mug of tea to a cold bottle of soda. Flexible thermal VR/AR devices are expected to enhance the level of artificial immersion and accurately portray VR/AR experiences \citep{Lee2021}.

\subsection{Energy harvesting applications}

The NaNbO$_{3}$ (NN) seeds are used
to texture a lead-free (K,Na)NbO$_{3}$-based piezoelectric ceramic
along the (001) direction. The chemical formula is 0.96(K$_{0.5}$Na$_{0.5}$)(Nb$_{0.965}$Sb$_{0.035}$)O$_{3}$-0.01CaZrO$_{3}$-0.03(Bi$_{0.5}$K$_{0.5}$) HfO$_{3}$
(KNN) at room temperature provides an outstanding combination of electromechanical
coefficients. At room temperature, the textured composition with a
5 wt\% NN template (KNN-5NN) has significantly better electromechanical
coefficients, d$_{33}\sim$ 590 pC/N, k$_{31}\sim$ 0.46, and
d$_{31}\sim$ 215$\times$10$^{-12}$C/N. Utilizing the textured KNN-5NN
ceramic, a flexible piezoelectric energy harvester (F-PEH) is created
and put to the test with cyclic force. In the off-resonance frequency
domain, F-PEH displays improved output voltage (V$_{oc}\sim$ 25V),
current (I$\sim$ 0.4A), and power density (P$_{D}\sim$ 5.5mW/m$^{2}$)
(R$_{L}$ of 10M). The textured F-PEH substantially exceeded energy
harvesting performance in comparison to the random ceramic KNN-0NN-based
F-PEH (V$_{oc}\sim$ 8V and I$\sim$ 0.1A), thanks to the high figure-of-merit
value (d$_{31}\times$ g$_{31}$) 3354$\times$10$^{-15}$m$^{3}/$J. This
study shows a method for texturing lead-free materials, and it is
further suggested that this technology be used for flexible energy
harvesting devices and sensors \citep{Purusothaman2023}. 

By using a straightforward, inexpensive, and large-area spin-casting/bar
coating technique, Park et al. \citep{Park2012} created a nanocomposite-based
nano-generator that makes use of piezoelectric BaTiO$_{3}$ nanoparticles
and general graphitic carbon (carbon nanotube and reduced graphene
oxide). A piezoelectric nanocomposite was produced by dispersing piezoelectric
BaTiO$_{3}$ nanoparticles and graphitic carbons in polydimethylsiloxane
(PDMS) elastomer. Piezoelectric materials were strengthened by graphitic
carbons' ability to operate as a stress agent, a dispersion to prevent
the precipitation of BaTiO$_{3}$ nanoparticles, and an electrical
nano-bridge to provide a conduction pathway in the polymer matrix.
A switching polarity test was run to ensure that the NCG sample generated
just the measured output signal. During the many bending and unbending
cycles, the NCG gadget regularly produced an open circuit voltage
(V$_{oc}$) of 3.2 V and a short-circuit current (I$_{sc}$) signal
of 250\textendash 350 nA.

Lin et al. \citep{Lin2012} used a simple, low-cost hydrothermal approach
to create bio-eco-compatible nanocomposite-based nano-generators based
on BaTiO$_{3}$ nanotubes. The piezoelectric nanocomposite was created
using BaTiO$_{3}$ nanotubes containing a polydimethylsiloxane (PDMS)
polymer and a simple poling method. A flexible and transparent nano-generator
was effectively built. The periodic bending and unbending of the top
and bottom electrodes established the electric potential at the electrode.
The created forward and reversed connection devices produced an output
voltage and current of roughly 6 V and 360 nA, respectively.

Further, Shin et al. \citep{Shin2014} describe highly efficient and
flexible nano-generators (NGs) made of hemispherically aggregated
BaTiO$_{3}$ NPs (BTO) and polyvinylidene fluoride-co-hexa-fluoro-propylene
(PVDF-HFP). To increase piezoelectric power generation, hemispherical
(BTO-PVDF-HFP) clusters were synthesized using solvent evaporation.
When stress is applied to the surface generally, the manufactured
NGs exhibit high electrical output up to 75 V and 15 A.

The above passage emphasizes the various uses of lead-free piezoelectric
materials and methods to improve the performance of piezoelectric
devices. Nevertheless, all of the aforementioned advancements are
grounded in empirical discoveries that offer valuable insights for
the progress of piezoelectric materials. The next part is dedicated
to exploring various mathematical models that can be used to advance
the development of lead-free piezoelectric materials. 

\subsection{Nano-electromechanical systems (NEMS)}

A self-powered continuous arterial-pulse monitoring system can improve the treatment quality, and hence, lengthen patient’s life. Conventional pulse monitoring sensors have significant power consumption drawbacks that limit their continuous operation. To overcome this limitation, a lead-free, biocompatible, and high-performance aluminum-nitride (AlN) piezoelectric nanogenerator-based self-powered (energy-harvesting) MEMS sensor is proposed which is capable of continuously monitoring arterial-pulse. This article presents a comprehensive physics-based mathematical modelling and numerical simulation results that deliver optimized design parameters for a novel piezoelectric thin-film MEMS diaphragm energy-converter (transducer) for improved sensitivity and fatigue stress life-span. The sensor yields a nano-generation sensitivity of 0.13V/kPa, which is much better than recently reported nanogenerator-based arterial-pulse sensors. Most materials encounter premature failure due to fatigue when subjected to cyclic loading. Damage-accumulation and cycles to fatigue-failure were also estimated using FEM simulations. The sensor’s viability is around 108.03 cycles before structural-failure for pressure amplitude-range from 10.6kPa to 16kPa (Range of arterial blood pressure) \citep{Dalin2022}.

Furthermore, in a novel approach, computational and experimental approaches establish microstructure design criteria and optimize rhombohedral Bi$_{0.5}$Na$_{0.4}$K$_{0.1}$TiO$_{3}$ from untextured (1 MRD), d$_{33}$ = 155 pC/N, to textured (4.41 MRDs), d$_{33}$ = 227pC/N. Two-dimensional orientation maps from electron backscatter diffraction on sequential parallel layers are used to computationally reconstruct three-dimensional samples, simulate local piezoelectric grain interactions, and demonstrate that lead-free piezoelectric microstructures can be improved in crystallographic and polarization texture. The approach uses single-crystal phase anisotropy to determine material anisotropic microstructure parameters that improve piezoelectric performance and reliability, linking structure and macroscopic length scales \citep{Lee2013}. Moreover, BiFeO$_{3}$-SrTiO$_{3}$ (BF-ST) ceramics are a unique lead-free dielectric material with high dielectric constants and thermal stability. To maximize piezoelectric performance, $(1-x)$BF-$x$ST $(0.32 \leq x \leq 0.44)$ ceramics along the morphotropic phase boundary is manufactured and carefully examined their microstructure and electrical characteristics. As ST concentration increases, the rhombohedral phase fraction decreases and the cubic phase fraction increases. At $x$ = 0.38, the grain size reaches 5.66 $\mu$m, with a heterogeneous core-shell architecture, strong remanent polarization (51.2 $\mu$C/cm$^{2}$), and a maximum d$_{33}$ value of 72 pC/N. In addition, impedance spectroscopy shows a conductive core and nonconductive shell in the sample. These results show that improved BF-ST ceramics may replace lead-based piezoelectric materials with excellent ferroelectric and piezoelectric characteristics \citep{Wang2024}.

Templated and reactive-templated grain formation is a practical technique for fabricating ceramics with a desired texture. This process involves sintering green compacts that consist of template grains arranged in a certain alignment, along with matrix grains that are randomly orientated. To get highly textured materials, it is necessary to remove the matrix grains. It is crucial to comprehend the growth behaviour of both template and matrix grains in order to establish the appropriate conditions for effectively eliminating the matrix grains \citep{KIMURA2016}. The ceramics' ferroelectric characteristics, as evidenced by PE loops, indicate that the surplus of Bi$_{2}$O$_{3}$ enhances the leakage conductivity while simultaneously stabilizing the P$_{r}$ and E$_{C}$. The ceramics exhibit optimal piezoelectric capabilities at $x$ = 0.01, with a piezoelectric coefficient of around 217 pC/N and a dielectric constant of approximately 243 pm/V. However, they demonstrate a significant sensitivity to changes in temperature due to formation of a desired texture. \citep{Qin2023}. Moreover, stannum-doped BaTiO$_{3}$ ceramics are prepared to be environmentally friendly and sustainable. These ceramics are used to attain a very high piezoelectric coefficient, d$_{33}$, of over 1100 pC/N. This is the highest value yet recorded for lead-free piezoceramics. The method and paradigm of the exceptional piezoelectricity established in this study offer a viable possibility for substituting lead-based piezoelectrics with lead-free alternatives \citep{Wang2020a}.

\subsection{Photostrictive and piezo-phototronics applications}
The amalgamation of the photovoltaic effect with the converse piezoelectric effect results in the occurrence of a photostrictive phenomenon \citep{Singh2022,Sun2007}. The direct conversion of optical energy to mechanical strain gives it potential in wireless optomechanical applications such light-driven actuators/sensors, solar energy harvesting, photoacoustic devices, and light-controlled smart devices \citep{Bartasyte2023,Uchino2017,Chen2021,Kundys2015}. As micro-electromechanical devices advance in the 21st century, photostrictive devices may offer an alternative to traditional units based on piezoelectric or magnetostrictive materials. A wireless, remote-controlled micro-electromechanical system using photostrictive effect can reduce weight without the need for hard connections or sophisticated electronics.

Organic-inorganic hybrid perovskites have emerged as a promising photovoltaic solar cell material in recent years \citep{Snaith2013,Green2014,Graetzel2014}. The high optoelectric performance of hybrid perovskites has led to their widespread use in various domains, including light-emitting diodes, photodetectors, and lasers \citep{Zhao2016,Tan2014,Dou2014,Xing2014} The photostrictive behaviour of halide perovskites has recently garnered attention from researchers. 
Wang and colleagues found a significant photostriction ($1.25\times10^{-3}$) in methylammonium lead iodide (MAPbI$_{3}$) using AFM \citep{Zhou2016}. Under 100 mW cm$^{-2}$ white-light illumination, MAPbI$_{3}$ crystal showed reproducible photostriction ($5\times10^{-5}$) in various directions. He and associates found that methylammonium lead bromide (MAPbBr$_{3}$) shown superior stability than MAPbI$_{3}$ in humid environments during photostriction \citep{Wei2017}. Recently proposed completely inorganic halide perovskites, such as CsPbX$_{3}$ ($X =$ Cl, Br, and I), offer superior stability than hybrid perovskites in some areas \citep{Nam2017,Protesescu2015,Sutton2016}. Singh et al. \citep{Singh2022} photostrictive composite comprising a photovoltaic polymer (PTB7-Th) as the matrix and piezoelectric material fibres (PMN-35PT). The suggested synthetic photostrictive composite has the ability to substitute lead-based photostrictive material found in nature. This not only creates opportunities for new uses but also allows for customization of the required characteristics. However, the main constituents of these hybrid perovskites is lead which is a toxic material and needed to be replaced by a lead-free alternative.

Many transition metal oxides with noncentral symmetry are part of the ferroelectric/piezoelectric family \citep{Saito2004,Haertling1999,Benedek2013}, and their photostrictive impact has been studied. Nonpolar perovskite oxides demonstrated significant photostrictive response, even when irradiated by visible light \citep{KorffSchmising2007,KorffSchmising2008,Cao2023,Liu2012}. The authors found a significant photostriction (1.12 \%) in SrRuO$_{3}$ (SRO) thin films, estimated using lattice variation calculations using power-dependent Raman mode shifts \citep{Wei2017a}. Chu and collaborators demonstrated a greater photostriction (2 \%) in epitaxial SrIrO$_{3}$ (SIO) film, with a response time estimated at a few picoseconds \citep{Yang2018}.

BiFeO$_{3}$, a leading multiferroic, displays ferroelectric and antiferromagnetic characteristics at ambient temperature \citep{Wang2003,Catalan2009}. The bulk photovoltaic effect of BiFeO$_{3}$ has been extensively researched in recent decades \citep{nechache2015bandgap,alexe2011tip,yang2010,choi2009switchable}. Due of its complex structure, ferroelectricity, magnetism, and light coupling, BiFeO$_{3}$'s photostrictive effect has garnered attention in recent years \citep{kundys2010light,kundys2012wavelength,lejman2014giant,schick2014localized,burkert2016optical}. Kundys et al. originally observed this phenomenon in BiFeO$_{3}$ crystals using a capacitance dilatometer \citep{kundys2010light}. The photostrictive response was seen using light sources from 365 to 940 nm \citep{kundys2012wavelength}. A maximal photostriction of $3.5\times10^{-5}$ was observed under 365 nm UV light at 326 Wm$^{-2}$. When irradiated by visible light (455-940 nm), the photostrictive response considerably decreased. The response time varies with wavelength, from 1.1 to 2.7 s, from 365 to 940 nm. This is faster than polymer-based photostrictive materials but not considerably better than other lead-based ferroelectrics like PLZT. Further research is needed to understand the relationship between light-induced lattice distortion and the macroscopic photostrictive performance of BiFeO$_{3}$, particularly for polycrystalline. Due to its ferroic nature, it is important to analyze the potential connection between lattice and spin degrees of freedom. Recent investigations demonstrated optical-controlled magnetic states in BiFeO$_{3}$ heterostructures using photostriction \citep{iurchuk2016optical,aftabi2018optically,aftabi2018magnetic}.

Moreover, recent research by Rubio-Marcos et al. demonstrates the photoinduced structural evolution of BaTiO$_{3}$ single crystals employing improved characterisation techniques \citep{rubio2015ferroelectric,rubio2018reversible,rubio2019photo}. Using atomic force microscopy, they detected a significant surface height shift in BaTiO$_{3}$ crystals when exposed to 532 nm light \citep{rubio2019photo}. Additionally, the height change was linearly dependent on light power, with a maximum photoinduced strain of 0.04 \% at 60 mW, equivalent to PLZT. Futher, Rubio-Marcos et al. \citep{RubioMarcos2023} suggest a potential method to develop a new generation of micro-scale photo-controlled ferroelectric-based devices and support the existence of a direct connection between optical absorbance and light-induced strain in polycrystalline BaTiO$_{3}$. They reported that the photo-response is directly correlated with light absorption at charged domain walls (CDWs) and the photoinduced strain increases as the thickness of the sample decreases, indicating that this phenomenon is more significant in thick films rather than bulk materials. The relationship between the photovoltaic and photostrictive effects and the surface properties of the sample is clearly apparent. The surface features play a crucial role in the production of high-efficiency photostrictive devices. The sample thickness and roughness are anticipated to play a vital role in the design of thick film bimorphs, aiming to improve the efficiency of a micromechanical device \citep{Poosanaas2000}.
Furthermore, the research on the photostrictive impact of (K$_{0.5}$Na$_{0.5}$)NbO$_{3}$ (KNN) systems, the most common lead-free ferroelectric material, is crucial. Fu et al. investigated KNN ceramics' photostriction and optimized performance with Ba(Ni$_{0.5}$Nb$_{0.5}$)O$_{3}$ (BNNO) \citep{fu2020boosting}. A photoinduced strain of 0.045 \% was observed for the 0.98KNN-0.02BNNO composition under 405 nm light at 30 W cm$^{-2}$. The response time to attain maximum strain was 270 s, longer than previous ferroelectric photostrictive materials. This suggests that photoinduced thermal expansion cannot be neglected. More research is needed to determine the photostrictive performance of KNN-based ferroelectric materials.

Moreover, the classic solid-state reaction approach was used to generate lead-free ferroelectric solid solutions with the composition (1-x)Bi$_{0.5}$(Na$_{0.77}$K$_{0.18}$Li$_{0.05}$)0.5TiO$_{3}-x$Sr(Nb$_{0.5}$Ni$_{0.5}$)O$_{3}$ ($x =0-0.05$) that showed exceptional photostriction of visible light response The bandgap of ferroelectric solid solutions reduces with Sr(Nb$_{0.5}$Ni$_{0.5}$)O$_{3}$ (SNN) doping, decreasing interaction between A cation and BO$_{6}$ octahedron. Light absorption efficiency increases with bandgap decrease. Further, the weakening of A-O bond contacts and considerable relaxation behaviour suggest weaker lattice restrictions. BO$_{6}$ octahedrons are more easily distorted by photogenerated carriers, promoting photostriction. The 0.95Bi$_{0.5}$(Na$_{0.77}$K$_{0.18}$Li$_{0.05}$)0.5TiO$_{3}$-0.05Sr(Nb$_{0.5}$Ni$_{0.5}$)O$_{3}$ sample exhibits impressive photoinduced strain ($10.58\times10^{-4}$), strong photostrictive efficiency ($10.58\times10^{-12} m^{3} W^{-1}$), and quick response time (10 s) \citep{Ren2022}. Further, Ren et al. \citep{Ren2023} examine the impact of Bi non-stoichiometry on the concentration of oxygen vacancies and the transition temperature from a ferroelectric to a relaxor phase in the material 0.95Bi$_{0.52}$NKT-0.05BNT. This, in turn, influences the photostrictive characteristics of the material. The material has a low band gap of 2.6 eV, a significant photostriction of 1.8 × 10$^{-3}$ at 405 nm and 42.4 kW/m$^{2}$, a high photostriction efficiency of 1.7 × 10$^{-11}$ m$^{3}$/W, and a rapid reaction time of 5-6 seconds. Laser-powered Raman spectra demonstrate the tiny deformation of TiO$_{6}$ octahedra caused by light, as well as the light-induced phase change in 0.95Bi$_{0.52}$NKT-0.05BNT. This makes it a promising choice for future opto-electromechanical systems. Furthermore, it should be noted that there is a need for inventive designs of devices that employ narrow band gap piezoelectrics for emerging applications. Moreover, the synergistic utilization of photovoltaic and piezoelectric phenomena will inspire unprecedented applications in the future that are currently unimaginable \citep{Bai2021}. Fang et al. \citep{Fang2021} further established BNT as a versatile substance, but also introduces a fresh prospect for practical application in photo-driven sensors and actuators. The study found that the BNT ceramic exhibits a relationship between the intensity of light and elongation, with a photostriction $\Delta$L$/$L of around 0.08 \% seen when the BNT was subjected to 405 nm laser light with a power of 20 kW/m$^{2}$. The average duration for the ascent or degradation of any object, around 6 seconds, is considerably swifter than the comparable timeframe for organic polymers.

Photostrictive performance is often better in materials with perovskite structure, whether inorganic or hybrid. The universal nature of photostrictive action is similar to electrostriction, which is theoretically present in all materials \citep{Li2014,zhang1998giant,zhang2002all,devonshire1954theory}. Electromechanical applications, like actuators and transducers, limit the use of materials with small fundamental electrostriction until ferroelectric/piezoelectric materials were discovered. These materials exhibit large strain due to a combination of electrostriction, piezoresponse, and ferroelectric/ferroelastic domain switching \citep{viola2013contribution,hao2019progress}. To study materials with strong optomechanical coupling, various factors must be considered. Ferroelectrics' large photostriction may not be solely due to photovoltaic or depolarization field screening effects. Future research should consider the contributions of electron-phonon coupling, photoelastic effect, and light-induced domain evolutions. While photostriction is commonly defined as light-induced nonthermal deformation, the photothermal effect can also cause thermal deformation during optomechanical coupling, which can be significant in some applications, particularly for organic materials \citep{lee2012photochemical,gelebart2017mastering,kohlmeyer2013wavelength}.

The photostrictive effect is a realistic method for wireless and remote control of micromechanical systems by directly converting light to mechanical strain.
Using light as power sources, it has the ability to replace or integrate inverse piezoelectric effect in electric-driven mechanical applications such actuators, motors, and acoustic devices. To eliminate electrical connections or electromagnetic noise in particular settings like vacuum and liquids, the photostrictive effect may be the best solution. The photostrictive effect is understudied compared to the piezoelectric effect, with no practical applications yet realized. It is important to note that the photostrictive effect is not the only interaction between light and matter, but also links other phenomena including photovoltaic, photolysis, and photocatalysis \citep{Bartasyte2023,Uchino2017,Chen2021}. Understanding the relationships between these occurrences will lead to novel concepts and applications that are underutilized. Recent studies have shown that light-induced lattice expansion enhances the efficiency and stability of perovskite solar cells. Further emphasis might be placed on the study of correlated electron physics that extends beyond the realm of light-matter interactions. Several photostrictive materials have been found to exhibit multifunctionality, wherein they demonstrate the simultaneous occurrence of at least two distinct effects. Greater emphasis must be placed on exploring the potential connections between optical, mechanical, electrical, magnetic, and other related factors.

\section{Theoretical developments and computational models for lead-free piezoelectric materials}
\markright{Theoretical developments and computational models for lead-free piezoelectric materials \hfill \empty}

\noindent Most of the above discussed literature focused on experimental
development of lead-free piezoelectric materials, which is quite useful
for the development of the piezoelectric field, however, it is not
a cost effective way but laid the foundation of numerical modelling
of piezoelectric materials. The mathematical models not only provide
flexibility to simulate experimental conditions but these also save
a lot of time and money simultaneously. The numerical models provide
a wide opportunity to conduct futuristic research which is difficult
to achieve by experimental methods. In this section, we are going
to explain the phenomenological models for electromechanical coupling,
thermoelectromechanical coupling, thermo-magneto-electromechanical
coupling, thermo-electric coupling, micro-scale phase-field thermo-electromechanical
coupling, macro-scale phase-field thermo-electromechanical couplig,
and viscoelastically damped thermo-electromechanical coupling for
piezoelectric materials.

\subsection{Electromechanical model}

The electromechanical model solely focuses on the piezoelectric effect, which involves the generation of electricity when a strain field is applied to the material, and vice versa. Additionally, there are several non-local or non-linear phenomena, such as flexoelectric effects (the generation of electricity due to strain gradients and vice versa) \citep{Yudin2013, JAGDISH2024}, and electrostrictive effects (the non-linear relationship between strain and electric fields) \citep{Sundar1992}. The model incorporates all of these impacts by taking into account the Gibbs free energy function \citep{He2019,BahramiSamani2010,Abdollahi2014,Kuang2009}, and demonstrated as:
\begin{eqnarray}
\phi(\pmb{\varepsilon},\pmb{\overrightarrow{E}}) & = & \frac{1}{2}\pmb{C}\pmb{\varepsilon}:\pmb{\varepsilon}-\pmb{e}\pmb{\overrightarrow{E}}\pmb{\varepsilon}-\frac{1}{2}\pmb{\overrightarrow{E}}^{T}\pmb{\epsilon}\pmb{\overrightarrow{E}}-\pmb{\mu}\pmb{\overrightarrow{E}}\nabla\pmb{\varepsilon}-\frac{1}{2}\pmb{\overrightarrow{E}}^{T}\pmb{G}\pmb{\varepsilon}\pmb{\overrightarrow{E}}.\label{eq:EMC}
\end{eqnarray}
The terms $\pmb{C}$, $\pmb{\epsilon}$, $\pmb{e}$, $\pmb{G}$,
and $\pmb{\mu}$ represent the elastic, permittivity, piezoelectric,
electrostrictive, and flexoelectric coefficients, respectively. In addition, the variables in the field include the strain tensor, represented as $\pmb{\varepsilon}$, the electric field, represented as $\pmb{\overrightarrow{E}}$, and the strain-gradient, represented as $\nabla\pmb{\varepsilon}$. The linear piezoelectric models often discussed in literature only take into account the first three components on the right-hand side, hence disregarding the nonlocal and nonlinear effects \citep{Prabhakar2013}.
The phenomenological relations that describe the behavior of the composites
are obtained by deriving them from Eq. (\ref{eq:EMC}).
\begin{eqnarray}
\pmb{\sigma} & = & \phi_{\pmb{\varepsilon}}=\pmb{C}\pmb{\varepsilon}-\pmb{e}\pmb{\overrightarrow{E}}-\frac{1}{2}\pmb{G}\pmb{\overrightarrow{E}}^{T}\pmb{\overrightarrow{E}},\label{eq:EMCsigma}
\end{eqnarray}
\begin{eqnarray}
\pmb{\hat{\sigma}} &= &  -\phi_{\nabla\pmb{\varepsilon}}=\pmb{\mu}\pmb{\overrightarrow{E}},\label{eq:Eq3}
\end{eqnarray}
\begin{eqnarray}
\pmb{\overrightarrow{D}} & = & -\phi_{\pmb{\overrightarrow{E}}}=\pmb{\epsilon}\pmb{\overrightarrow{E}}+\pmb{e}\pmb{\varepsilon}+\pmb{\mu}\nabla\pmb{\varepsilon},\label{eq:EMC electric disp}
\end{eqnarray}
here, $\pmb{\sigma}$ denotes the stress tensor, $\pmb{\hat{\sigma}}$ denotes the higher order stress tensor, and $\pmb{\overrightarrow{D}}$ denotes the electric flux density vector. The constitutive relations are also susceptible to the governing equilibrium rules
 \citep{Sharma2010,Mao2014}:
\begin{eqnarray}
\nabla\cdot(\pmb{\sigma}-\pmb{\hat{\sigma}})+\pmb{\overrightarrow{F}} & = & 0,\label{eq:Eq6}
\end{eqnarray}
\begin{eqnarray}
\nabla\cdot\pmb{\overrightarrow{D}} & = & 0.\label{eq:Eq7}
\end{eqnarray}
In this numerical model, the body force vector $\pmb{\overrightarrow{F}}$ is assumed to be missing. The equations of motion, namely Eqs. (\ref{eq:Eq6}) and (\ref{eq:Eq7}), are solved using Finite Element Analysis. This analysis considers the phenomenological connections described by Eqs. (\ref{eq:EMC}) and (\ref{eq:EMCsigma}). If the higher-order stress components $\pmb{\hat{\sigma}}$ are not present, Equation (\ref{eq:Eq6}) would simplify to the linear momentum balance in the classical elastic formalism. The higher-order stresses can be interpreted as stresses related to moments, and the balance equation can be seen as including the balance for both linear and rotational momentum. The gradient equations express the relationships between linear strain and mechanical displacement, as well as between electric field and electric potential change. The following statements are interconnected:
\begin{eqnarray}
\pmb{\varepsilon} & = & \frac{1}{2}(\nabla \pmb{\overrightarrow{u}}+\nabla \pmb{\overrightarrow{u}}^{T}),\label{eq:Eq9}
\end{eqnarray}
\begin{eqnarray}
\pmb{\overrightarrow{E}} & = & -\nabla V,\label{eq:Eq10}
\end{eqnarray}
where $\pmb{\varepsilon}$, $\pmb{\overrightarrow{E}}$, $\pmb{\overrightarrow{u}}$, and $V$ are the strain tensor, electric field vector, mechanical displacement
vector, and electric potential respectively.

\subsection{Thermo-electromechanical model}

Piezoelectric materials have distinct behaviour when subjected to a thermal field, resulting in the formation of thermal stress and strain in the mechanical field. Additionally, they display an opposing thermoelectric effect in contrast to the piezoelectric effect. This approach is particularly advantageous for accurately forecasting the behaviour of sensors and actuators in high temperature and cryogenic applications, where fluctuations in temperature significantly impact the performance of the piezoelectric material. Thus, it is necessary to modify the Gibbs energy function. In the linear theory context, the Helmholtz free energy function $\phi$ for our system takes into account three independent variables $(\pmb{\varepsilon},\pmb{\overrightarrow{E}},\theta)$. The function is defined as follows \citep{Chandrasekharaiah1988, Akshayveer2024a}:
\begin{eqnarray}
\phi(\pmb{\varepsilon},\pmb{\overrightarrow{E}},\theta) & = & \frac{1}{2}\pmb{C}\pmb{\varepsilon}:\pmb{\varepsilon}-\pmb{e}\pmb{\overrightarrow{E}}\pmb{\varepsilon}-\frac{1}{2}\pmb{\overrightarrow{E}}^{T}\pmb{\epsilon}\pmb{\overrightarrow{E}}-\pmb{\mu}\pmb{\overrightarrow{E}}\nabla\pmb{\varepsilon}-\pmb{\beta}\theta\pmb{\varepsilon}-\pmb{\eta}\pmb{\overrightarrow{E}}\theta-\frac{1}{2}a_{\theta}\theta^{2}.\label{eq:Eq1}
\end{eqnarray}
The variables $\pmb{C}$ represent the elastic constants, $\pmb{\epsilon}$ represent the dielectric constants, $\pmb{e}$ represent the piezoelectric constants, $\pmb{\mu}$ represent the flexoelectric constants, $\pmb{\beta}$ represent the thermal modulus, $\pmb{\eta}$ represents the thermoelectric constants, and $a_{\theta}$ represent the heat capacity coefficients. The heat capacity coefficients $a_{\theta}$ are calculated by dividing the heat capacity at constant volume $C_{\pmb{\varepsilon}}^{V}$ by the ambient temperature $\theta_{R}$. The heat capacity at constant volume refers to the ability of a substance to absorb heat without changing its volume, while the ambient temperature is the temperature of the surrounding environment. The constitutive relationships for thermo-electro-mechanical coupling are obtained from Equation (\ref{eq:Eq1}) and may be expressed as
 \citep{Patil2009}:
\begin{eqnarray}
\pmb{\sigma} & = & \phi_{\pmb{\varepsilon}}=\pmb{C}\pmb{\varepsilon}-\pmb{e}\pmb{\overrightarrow{E}}-\pmb{\beta}_\theta,\label{eq:Eq2}
\end{eqnarray}
\begin{eqnarray}
\pmb{\overrightarrow{D}} & = & -\phi_{\pmb{\overrightarrow{E}}}=\pmb{\epsilon}\pmb{\overrightarrow{E}}+\pmb{e}\pmb{\varepsilon}+\pmb{\eta}\theta-\pmb{\mu}\nabla\pmb{\varepsilon},\label{eq:Eq4}
\end{eqnarray}
\begin{eqnarray}
S= &  \phi_{\theta} & =\pmb{\beta}\pmb{\varepsilon}+\pmb{\eta}\pmb{\overrightarrow{E}}+a_{\theta}\theta,\label{eq:Eq5}
\end{eqnarray}
where, the symbol $S$ represents the entropy, whereas $\theta$ represents the temperature difference.
The symbols $\pmb{\sigma}$ and $\pmb{\varepsilon}$ represent the elastic stress and strain tensor, respectively. The vector $\pmb{\overrightarrow{D}}$ represents the electric flux density, while $\pmb{\overrightarrow{E}}$ represents the electric field. Additionally, the equations that establish equilibrium in the mechanical and electrical fields are represented by Eqs. (\ref{eq:Eq6}) and (\ref{eq:Eq7}), respectively. On the other hand, the equation regulating the thermal field is as follows \citep
{Krishnaswamy2019,Patil2009}:
\begin{eqnarray}
\nabla\cdot\pmb{\overrightarrow{Q}}+Q_{gen} & = & 0,\label{eq:Eq8}
\end{eqnarray}
where Eqn. (\ref{eq:Eq8}) simplifies to a steady-state heat conduction equation in the absence of the heat production term $Q_{gen}$.
The gradient equations represent the connections between linear strain and mechanical displacement, whereas the equations Eqs. (\ref{eq:Eq9}) and (\ref{eq:Eq10}) describe the link between electric field and electric potential  \citep{Krishnaswamy2020a,Patil2009}:

\begin{equation}
\pmb{\overrightarrow{Q}}=-\pmb{k}\nabla\theta,\label{eq:Eq11}
\end{equation}
where $\pmb{\overrightarrow{Q}}$, $\pmb{k}$, and $\theta$ are the thermal
field vector, thermal conductivity and temperature change from the
reference, respectively.

\subsection{Thermo-magneto-electromechanical model}

Lead-free piezoelectric material exhibits distinct responses to magnetic fields, thereby necessitating adjustments to the thermo-electromechanical model in order to accurately forecast its behaviour. The Helmholtz free energy function for the system, which incorporates the magnetic field into the thermo-electromechanical model, may be expressed as:
\begin{eqnarray}
\phi(\pmb{\varepsilon},\pmb{\overrightarrow{E}},\theta,\pmb{\overrightarrow{B}}) & = & \phi(\pmb{\varepsilon},\pmb{\overrightarrow{E}},\theta)+\frac{1}{2}\pmb{\nu}\pmb{\overrightarrow{B}}^{2}-\pmb{\pi}\pmb{\varepsilon}\pmb{\overrightarrow{B}}-\pmb{\lambda}\pmb{\overrightarrow{E}}\pmb{\overrightarrow{B}}-\pmb{\tau}\pmb{\overrightarrow{B}}\theta,\label{eq:Mag-GE}
\end{eqnarray}
The derived constitutive equations in this instance exhibit the subsequent
structure:
\begin{eqnarray}
\pmb{\sigma} & = & \phi_{\pmb{\varepsilon}}=\pmb{C}\pmb{\varepsilon}-\pmb{e}\pmb{\overrightarrow{E}}-\pmb{\beta}\theta-\pmb{\pi}\pmb{\overrightarrow{B}},\label{eq:Mag-sigma}
\end{eqnarray}
\begin{eqnarray}
\pmb{\overrightarrow{D}} & = & - \phi_{\pmb{\overrightarrow{E}}}=\pmb{e}\pmb{\varepsilon}+\pmb{\epsilon}\pmb{\overrightarrow{E}}+\pmb{\eta}\theta+\pmb{\lambda}\pmb{\overrightarrow{B}},\label{eq:Mag_ED}
\end{eqnarray}
\begin{eqnarray}
\pmb{\overrightarrow{H}} & = & \phi_{\pmb{\overrightarrow{B}}}=-\pmb{\pi}\pmb{\varepsilon}-\pmb{\lambda}\pmb{\overrightarrow{E}}-\pmb{\tau}\theta+\pmb{\nu}\pmb{\overrightarrow{E}},\label{eq:Mag_H}
\end{eqnarray}
\begin{eqnarray}
S & = & \phi_{\theta}=\pmb{\beta}\pmb{\varepsilon}+\pmb{\eta}\pmb{\overrightarrow{E}}+a_{\theta}\theta+\pmb{\tau}\pmb{\overrightarrow{B}}.\label{eq:Mag_HE}
\end{eqnarray}
The symbols $\pmb{\pi}$, $\pmb{\lambda}$, and $\pmb{\tau}$ denote the coupling constants associated with the piezo-magnetic, electro-magnetic, and thermo-magnetic effects, respectively. The sign $\pmb{\nu}$ is equivalent to the reciprocal of $\mu$, where $\mu$ indicates permeability. The study conducted by Patil (2009) utilizes this model to examine the thermo-magneto-electro-elastic processes in nanowire superlattices. A new advancement in \citep{Willatzen2006,Melnik2007} has provided a novel generalization that takes into account the impact of nonlinear strain effects. The model necessitates an extra governing equation for the magnetic field strength $\pmb{\overrightarrow{H}}$, which is expressed as follows:
\begin{eqnarray}
\nabla\times \pmb{\overrightarrow{H}} & = & 0\label{eq:mag-GE}
\end{eqnarray}

\subsection{Thermo-electrical model}

Only thermoelectric constitutive equations are necessary for modeling
when the thermo-electric effects greatly outweigh other effects like
as piezoelectricity, flexoelectricity, and ferroelectricity. In this
scenario, constitutive equations are often expressed in terms of the
electric current density ($\pmb{\overrightarrow{j}}^{el}$) and the heat flux ($\pmb{\overrightarrow{Q}}$);
\begin{eqnarray}
\pmb{\overrightarrow{j}}^{el} & = & \rho_{e} \pmb{\overrightarrow{E}}-\rho_{e} S_{b}\nabla\theta,\label{eq:TE1}
\end{eqnarray}
\begin{eqnarray}
\pmb{\overrightarrow{Q}} & = & -\rho_{e} S_{b}\theta \nabla V-(\pmb{k}+S_{b}^{2}\rho\theta)\nabla\theta,\label{eq:TE2}
\end{eqnarray}
The electric conductivity is denoted by $\rho_{e}$, the thermal conductivity
is denoted by $\pmb{k}$, and the Seebeck coefficient is denoted by
$S_{b}$. It is important to note that these coefficients can vary
with temperature. For isotropic materials, the thermoelectric figure
of merit,through these coefficient, is defined as $Z=\theta\rho S_{b}^{2}/\pmb{k}$
\citep{Bies2002}.The device application environment was the focus
of a study on linked thermoelectric models in reference \citep{Wang2012}.
Equation (\ref{eq:TE2}) extends Fourier's law to incorporate both
the Peltier and Thompson effects. The equations that govern this particular
scenario are:
\begin{eqnarray}
\nabla\cdot\pmb{\overrightarrow{j}}^{el} & = & 0\label{eq:GE-TE}
\end{eqnarray}
\begin{eqnarray}
\nabla\cdot\pmb{\overrightarrow{Q}}-K & = & 0\label{eq:GE-TE2}
\end{eqnarray}
with $K=-\pmb{\overrightarrow{j}}^{el}\cdot\nabla V$. The inclusion of the convective term
(latter term) in the final model renders it a nonlinear representation
that extends beyond traditional linear models. This includes a thermoelectric
model that incorporates the linear Seebeck effect \citep{PerezAparicio2007},
where $\pmb{\overrightarrow{Q}}=-\pmb{k}\nabla\theta$. In this case, the coupling is effectively
implemented through Equation (\ref{eq:GE-TE}).

\subsection{Microscale phase-field thermo-electromechanical model}

Piezoelectric materials not only demonstrate the piezoelectric and thermoelectric effects, but they also display ferroelectric polarization. This polarization is responsible for phase changes and microdomain switching when a heat field is applied. In this model, the phase transitions and ferroelectric domain transitions are accounted for by incorporating the Landau-Ginzburg-Devonshire contribution, (in \citep{Ahluwalia,Morozovska,Wang, Akshayveer2024, Akshayveer2024b}). This contribution is added to the Helmholtz free energy equation (Eq. (\ref{eq:free_energy})), which depends on the electric field vector, strain vector, polarization vector, gradient of polarization vector, and temperature ($\theta$) of the system \citep{Borrelli2019}:
\begin{eqnarray}
\phi(\pmb{\overrightarrow{E}},\pmb{\varepsilon}(\pmb{\overrightarrow{u}}),\pmb{\overrightarrow{p}},\pmb{\nabla\overrightarrow{p}},\theta) & = & \frac{1}{2}\pmb{\lambda}\left|\pmb{\nabla\overrightarrow{p}}\right|^{2}+W(\theta,\pmb{\overrightarrow{p}})+\frac{1}{2}\pmb{\varepsilon^{el}:C\varepsilon^{el}}-(\theta-\theta_{R})\pmb{\beta:\varepsilon}(\pmb{\overrightarrow{u}})-\frac{1}{2}\pmb{{\overrightarrow{E}}^{T}\epsilon{\overrightarrow{E}}}\nonumber \\
 & - &\pmb{\mu\overrightarrow{E}\nabla\varepsilon^{el}}-\pmb{e\overrightarrow{E}\varepsilon^{el}}-\pmb{\overrightarrow{p}\overrightarrow{E}}-\pmb{\eta}(\theta-\theta_{R})\pmb{\overrightarrow{E}}.\label{eq:free_energy}
\end{eqnarray}
The first component of Equation (\ref{eq:free_energy}) represents the Landau gradient energy, which arises from the spatial gradient of the polarization vector $\pmb{\overrightarrow{p}}$ ($[p_{1} \quad p_{3}]^{T}$). Within the two-dimensional domain, the second term represents the Landau-Ginzburg-Devonshire function of free energy (refer to Equation (\ref{eq:polynomial2D}) below). The other terms encompass elastic deformation energy, thermal deformation energy, dielectric energy, flexoelectric energy, piezoelectric energy, electric polarization energy, and thermoelectric energy, respectively. The variables $\pmb{C}$, $\pmb{\epsilon}$, $\pmb{e}$, $\pmb{\mu}$, $\pmb{\beta}$, indicate the elastic constants, dielectric constants, piezoelectric constants, flexoelectric constants, thermal expansion coefficient, and thermoelectric constants correspondingly. In addition, $\pmb{\overrightarrow{E}}$ represents the vector that represents the intensity of the electric field. $\pmb{\varepsilon(\overrightarrow{u})}$ is the tensor that represents the total mechanical strain, and $\pmb{\overrightarrow{u}}$ represents the vector that represents the mechanical displacement.
The symbol $\theta$ represents the temperature, whereas $\theta_{R}$ represents the reference temperature.
The total mechanical strain, denoted as $\pmb{\varepsilon(\overrightarrow{u})}$, is composed of two distinct components. The first component is the elastic strain tensor, $\pmb{\varepsilon^{el}}$, which is linearly related to the stress tensor, $\pmb{\sigma}$. The second component is the transformation strain tensor, $\pmb{\varepsilon^{t}(\overrightarrow{p})}$, which is associated with phase changes and ferroelectric transitions.
\begin{eqnarray}
\pmb{\varepsilon}(\pmb{\overrightarrow{u}}) & = & \frac{1}{2}(\pmb{\nabla\overrightarrow{u}+\nabla \overrightarrow{u}}^T)=\pmb{\varepsilon^{el}}+\pmb{\varepsilon^{t}}(\pmb{\overrightarrow{p}}),\label{eq:total_strain}
\end{eqnarray}
The variable $\pmb{\varepsilon^{t}(\overrightarrow{p})}$ represents the deformation of the crystal unit cell resulting from the ferroelectric transition. The tensor represents a deformation that happens in the same direction as the ferroelectric polarization vector $\pmb{\overrightarrow{p}}$. The deformation is exactly proportional to the modulus of the vector $\pmb{\overrightarrow{p}}$, with a material constant $\gamma$ as the proportionality factor \citep{Kamlah1999}. This relationship may be expressed as:
\begin{eqnarray}
\pmb{\varepsilon^{t}}(\pmb{\overrightarrow{p}}) & = & \gamma\left|\pmb{\overrightarrow{p}}\right|(\pmb{\overrightarrow{n}}\otimes\pmb{\overrightarrow{n}}-\frac{1}{d}\pmb{I}),\pmb{\overrightarrow{n}}\coloneqq\frac{\pmb{\overrightarrow{p}}}{\left|\pmb{\overrightarrow{p}}\right|}.\label{eq:transformation_strain}
\end{eqnarray}
For three-dimensional deformations, the value of $d$ is equal to 3 \citep{Kamlah2001}. When considering planar strain, where we assign $n_2 = 0$ and $\varepsilon_{22}=0$, the value of the factor $d$ should be chosen as $d=2$. To be more specific, the transformation strain refers to an elongation that is directly proportional to the product of the magnitude of the polarization vector, $|\pmb{\overrightarrow{p}}|$, and the value of $\gamma$. The component of the vector $\mathbf{\pmb{\overrightarrow{n}}}$ that is perpendicular to the plane undergoes a reduction in length, whereas the component that is not in the plane remains the same. This modification for $d$ guarantees that there will be no alteration in volume in plane strain circumstances. It is important to emphasize that the transformation strain demonstrates symmetry with respect to the unit vector $\pmb{\overrightarrow{n}}$, implying that the same deformation is associated with many polarization states.
Hence, domains that exhibit a mixture of positive and negative polarization states The vector $(\pm\pmb{\overrightarrow{n}})$ can have a polarization that reaches zero, even while it still possesses a non-zero transformation strain.
The tensor multiplication of the unit vector $\pmb{\overrightarrow{n}}$ with its transpose vector $\pmb{\overrightarrow{n}}\otimes\pmb{\overrightarrow{n}}$ results in a symmetric transformation strain. This symmetry with reference to $\pmb{\overrightarrow{n}}$ leads to opposing polarization states for the same deformation.
Consequently, when a fluctuating electric field is supplied, the strain exhibits a butterfly-shaped form and a hysteresis curve for the polarization vector \citep{Landis2002,McMeeking2002,liu2020}.

The Landau-Ginzburg-Devonshire free energy function $W^{3D}(\theta,\pmb{\overrightarrow{p}})$ describes the phase transition caused by temperature and also provides information on other thermal properties of the material, such as specific heat. According to Landau's suggestions, the Landau-Ginzburg-Devonshire free energy may be represented as a polynomial function of the order parameter $\pmb{\overrightarrow{p}}$, which must satisfy certain symmetry conditions depending on the material's symmetry. The Landau-Ginzburg-Devonshire free energy function, which represents a polynomial, is defined for a three-dimensional anisotropic crystal is defined as \citep{Voelker2011,Indergand2020,Indergand2021}:
\begin{eqnarray}
W^{3D}(\theta,\pmb{\overrightarrow{p}}) & = &\alpha_{1}\frac{\theta_{c}-\theta}{\theta_{c}}(p_{1}^{2}+p_{2}^{2}+p_{3}^{2})+\alpha_{11}(p_{1}^{4}+p_{2}^{4}+p_{3}^{4})+\alpha_{12}\frac{\theta_{c}-\theta}{\theta_{c}}(p_{1}^{2}p_{2}^{2}+p_{2}^{2}p_{3}^{2}+p_{3}^{2}p_{1}^{2})\nonumber \\
 & + &\alpha_{111}(p_{1}^{6}+p_{2}^{6}+p_{6}^{4})+ \alpha_{112}[p_{1}^{4}(p_{2}^{2}+p_{3}^{2})+p_{2}^{4}(p_{3}^{2}+p_{1}^{2})+p_{3}^{4}(p_{1}^{2}+p_{2}^{2})]+\alpha_{123}(p_{1}^{2}p_{2}^{2}p_{3}^{2}).\label{eq:polynomial}
\end{eqnarray}
The material properties are denoted by $\alpha_{1}$, $\alpha_{11}$, $\alpha_{12}$, $\alpha_{111}$, $\alpha_{112}$, and $\alpha_{123}$, whereas $\theta_{c}$ denotes the Curie temperature.
Devonshire \citep{Devonshire1949} focuses only on the term $\alpha_{111}(p_{1}^{6}+p_{2}^{6}+p_{3}^{6})$ in the sixth-order expansion due to the minimal influence of other sixth-order terms. The polynomial $W^{3D}(\theta,\pmb{\overrightarrow{p}})$ may be expressed in the following form:
\begin{eqnarray}
W^{3D}(\theta,\pmb{\overrightarrow{p}}) & = &\alpha_{1}\frac{\theta_{c}-\theta}{\theta_{c}}(p_{1}^{2}+p_{2}^{2}+p_{3}^{2})+\alpha_{11}(p_{1}^{4}+p_{2}^{4}+p_{3}^{4})+\alpha_{12}\frac{\theta_{c}-\theta}{\theta_{c}}(p_{1}^{2}p_{2}^{2}+p_{2}^{2}p_{3}^{2}+p_{3}^{2}p_{1}^{2})\nonumber \\
 & + &\alpha_{111}(p_{1}^{6}+p_{2}^{6}+p_{3}^{6}).\label{eq:polynomial2}
\end{eqnarray}
The minimization of the Helmholtz free energy function $\phi$ with respect to the vector $\pmb{\overrightarrow{p}}$ will also result in the minimization of $W^{3D}(\theta,\pmb{\overrightarrow{p}})$, which has a global minimum at $\pmb{\overrightarrow{p}}=0$ for every $\theta\geq\theta_{0}$ in the absence of an electric field. The symbol $\theta_{0}$ denotes the transition temperature, which is in close proximity to the Curie temperature $\theta_{c}$. 

Furthermore, it is hypothesized that the polarization component aligned with the translation axis has no significance ($p_{2} = 0$). In these cases, a perovskite ferroelectric that undergoes a transition from a cubic to a tetragonal phase can be regarded as an extra two-dimensional ferroelectric with a transition from a square to a rectangular phase, along with similar two-dimensional transitions in other phase changes \citep{Ahluwalia,Morozovska,Wang}. The polynomial $W^{2D}(\theta,\pmb{\overrightarrow{p}})$ is expressed for the current 2-dimensional $x_{1}-x_{3}$ computational domain:
\begin{eqnarray}
W^{2D}(\theta,\pmb{\overrightarrow{p}})  = \alpha_{1}\frac{\theta_{c}-\theta}{\theta_{c}}(p_{1}^{2}+p_{3}^{2})+\alpha_{11}(p_{1}^{4}+p_{3}^{4})+\alpha_{12}\frac{\theta_{c}-\theta}{\theta_{c}}p_{3}^{2}p_{1}^{2}+\alpha_{111}(p_{1}^{6}+p_{3}^{6}).\label{eq:polynomial2D}
\end{eqnarray}
In order to minimize $W^{3D}(\theta,\pmb{\overrightarrow{p}})$, we need to minimize the Helmholtz free energy function with respect to $\pmb{\overrightarrow{p}}$ and we will get:
\begin{equation}
\phi_{\pmb{\overrightarrow{p}}} = W^{3D}_{\pmb{\overrightarrow{p}}}(\theta,\pmb{\overrightarrow{p}})-\pmb{\overrightarrow{E}} = 0. \label{A1}
\end{equation}
Now all three components of electric field can be written as:
\begin{equation}
E_{1} = 2\alpha_{1}\frac{\theta_{c}-\theta}{\theta_{c}}p_{1}+4\alpha_{11}p_{1}^{3}+2\alpha_{12}\frac{\theta_{c}-\theta}{\theta_{c}}p_{1}(p_{2}^{2}+p_{3}^{2})+6\alpha_{111}p_{1}^{5}, \label{A2}
\end{equation}
\begin{equation}
E_{2} = 2\alpha_{1}\frac{\theta_{c}-\theta}{\theta_{c}}p_{2}+4\alpha_{11}p_{2}^{2}+2\alpha_{12}\frac{\theta_{c}-\theta}{\theta_{c}}p_{2}(p_{2}^{2}+p_{1}^{2})+6\alpha_{111}p_{2}^{5}, \label{A3}
\end{equation}
\begin{equation}
E_{3} = 2\alpha_{1}\frac{\theta_{c}-\theta}{\theta_{c}}p_{3}+4\alpha_{11}p_{3}^{2}+2\alpha_{12}\frac{\theta_{c}-\theta}{\theta_{c}}p_{3}(p_{1}^{2}+p_{2}^{2})+6\alpha_{111}p_{3}^{5}. \label{A4}
\end{equation}
In absence of electric field, the solution of the Eqs. (\ref{A2}), (\ref{A3}), and (\ref{A4}) can be written as:
\begin{equation}
p_{1} =0, \alpha_{1}\frac{\theta_{c}-\theta}{\theta_{c}}+2\alpha_{11}p_{1}^{2}+\alpha_{12}\frac{\theta_{c}-\theta}{\theta_{c}}(p_{2}^{2}+p_{3}^{2})+3\alpha_{111}p_{1}^{4} = 0, \label{A5}
\end{equation}
\begin{equation}
p_{2} =0, \alpha_{1}\frac{\theta_{c}-\theta}{\theta_{c}}+2\alpha_{11}p_{2}^{2}+\alpha_{12}\frac{\theta_{c}-\theta}{\theta_{c}}(p_{3}^{2}+p_{1}^{2})+3\alpha_{111}p_{2}^{4} = 0, \label{A6}
\end{equation}
\begin{equation}
p_{3} =0, \alpha_{1}\frac{\theta_{c}-\theta}{\theta_{c}}+2\alpha_{11}p_{3}^{2}+\alpha_{12}\frac{\theta_{c}-\theta}{\theta_{c}}(p_{1}^{2}+p_{2}^{2})+3\alpha_{111}p_{3}^{4} = 0. \label{A7}
\end{equation}
In order to describe the first-order phase transition, we use the assumption that $\alpha_{111}$ and $\alpha_{12}$ are positive, $\alpha_{11}$ is negative, and $\alpha_{1}\frac{\theta_{c}-\theta}{\theta_{c}}$ fluctuates with temperature, switching signs when the transition temperature is attained. When the value of $\alpha_{1}\frac{\theta_{c}-\theta}{\theta_{c}}$ is negative, the free energy will reach its lowest and lead to polarizations of a restricted magnitude. In equations (\ref{A5}), (\ref{A6}), and (\ref{A7}), the conditions $p_{1} =0$, $p_{2} =0$, and $p_{3} =0The symbol $ represents the solution above the Curie temperature ($\theta>\theta_{C}$) when the material demonstrates its paraelectric character and the polarization disappears in the absence of an electric field. The phase in question is often known as the cubic phase, and the solution for this phase may be expressed as:
\begin{eqnarray}
Cubic & : & p_{1}=p_{2}=p_{3}=0,\label{A8}
\end{eqnarray}
Below Curie temperature $\theta_{C}$, piezoelectric material may exhibit many phase structures, including tetragonal, orthorhombic, and rhombohedral, depending on the polarization direction. The polarization in the tetragonal phase is oriented in the (0,0,1) direction. Thus, the solution for the tetragonal phase is as follows:
\begin{eqnarray}
Tetragonal & : & p_{1}=p_{2}=0,p_{3}\neq0, \quad \alpha_{1}\frac{\theta_{c}-\theta}{\theta_{c}}+2\alpha_{11}p_{3}^{2}+3\alpha_{111}p_{3}^{4}=0.\label{A9}
\end{eqnarray}
The polarization direction in the orthorhombic phase is in the (0,1,1) direction, whereas in the rhombohedral phase it is along the (1,1,1) direction. Hence, the solution for the orthorhombic and rhombohedral phases may be expressed as:
\begin{eqnarray}
Orthorombic & : & p_{1}=0, p_{2}=p_{3}\neq0, \quad \alpha_{1}\frac{\theta_{c}-\theta}{\theta_{c}}+(2\alpha_{11}+\alpha_{12}\frac{\theta_{c}-\theta}{\theta_{c}})p_{3}^{2}+3\alpha_{111}p_{3}^{4}=0,\label{A10}
\end{eqnarray}
\begin{eqnarray}
Rhombohedral & : & p_{1}=p_{2}=p_{3}\neq0, \quad \alpha_{1}\frac{\theta_{c}-\theta}{\theta_{c}}+2(\alpha_{11}+\alpha_{12}\frac{\theta_{c}-\theta}{\theta_{c}})p_{3}^{2}+3\alpha_{111}p_{3}^{4}=0.\label{A11}
\end{eqnarray}
On the basis of above solution, the Landau-Ginzburg-Devonshire free energy function for each phase can be written as:
\begin{eqnarray}
Cubic & : & W^{3DC}(\theta,\pmb{\overrightarrow{p}})=0,\label{A12}
\end{eqnarray}
\begin{eqnarray}
Tetragonal & : & W^{3DT}(\theta,\pmb{\overrightarrow{p}})=\alpha_{1}\frac{\theta_{c}-\theta}{\theta_{c}}p_{3}^{2}+\alpha_{11}p_{3}^{4}+\alpha_{111}p_{3}^{6},\label{A13}
\end{eqnarray}
\begin{eqnarray}
Orthorombic & : & W^{3DO}(\theta,\pmb{\overrightarrow{p}})=2\alpha_{1}\frac{\theta_{c}-\theta}{\theta_{c}}p_{3}^{2}+(2\alpha_{11}+\alpha_{12}\frac{\theta_{c}-\theta}{\theta_{c}})p_{3}^{4}+2\alpha_{111}p_{3}^{6},\label{A14}
\end{eqnarray}
\begin{eqnarray}
Rhombohedral & : & W^{3DR}(\theta,\pmb{\overrightarrow{p}})=3\alpha_{1}\frac{\theta_{c}-\theta}{\theta_{c}}p_{3}^{2}+3(\alpha_{11}+\alpha_{12}\frac{\theta_{c}-\theta}{\theta_{c}})p_{3}^{4}+3\alpha_{111}p_{3}^{6}.\label{A15}
\end{eqnarray}

In a two-dimensional problem on the $x_{1}-x_{3}$ plane, the polarization vector has components $p_{1}$ and $p_{3}$ only. Hence, minimizing $\phi$ with regard to $\pmb{\overrightarrow{p}}$ will likewise result in the minimization of $W^{2D}(\theta,\pmb{\overrightarrow{p}})$.
\begin{equation}
\phi_{\pmb{\overrightarrow{p}}} = W^{2D}_{\pmb{\overrightarrow{p}}}(\theta,\pmb{\overrightarrow{p}})-\pmb{\overrightarrow{E}} = 0. \label{A16}
\end{equation}
The solution of Eq. (\ref{A16}) will give us both component of electric field along $x_{1}-x_{3}$, which can be written as:
\begin{equation}
E_{1} = 2\alpha_{1}\frac{\theta_{c}-\theta}{\theta_{c}}p_{1}+4\alpha_{11}p_{1}^{3}+2\alpha_{12}\frac{\theta_{c}-\theta}{\theta_{c}}p_{1}p_{3}^{2}+6\alpha_{111}p_{1}^{5}, \label{A17}
\end{equation}
\begin{equation}
E_{3} = 2\alpha_{1}\frac{\theta_{c}-\theta}{\theta_{c}}p_{3}+4\alpha_{11}p_{3}^{3}+2\alpha_{12}\frac{\theta_{c}-\theta}{\theta_{c}}p_{3}p_{1}^{2}+6\alpha_{111}p_{3}^{5}. \label{A18}
\end{equation}
In the absence of electric field, the solution of Eqs. \ref{A17}, and \ref{A18} can be written as:
\begin{equation}
p_{1} =0, \alpha_{1}\frac{\theta_{c}-\theta}{\theta_{c}}+2\alpha_{11}p_{1}^{2}+\alpha_{12}\frac{\theta_{c}-\theta}{\theta_{c}}p_{3}^{2}+3\alpha_{111}p_{1}^{4} = 0, \label{A19}
\end{equation}
\begin{equation}
p_{3} =0, \alpha_{1}\frac{\theta_{c}-\theta}{\theta_{c}}+2\alpha_{11}p_{3}^{2}+\alpha_{12}\frac{\theta_{c}-\theta}{\theta_{c}}p_{1}^{2}+3\alpha_{111}p_{3}^{4} = 0. \label{A20}
\end{equation}
Eqns. (\ref{A19}) and (\ref{A20}) are influenced by $p_{1} =0$ and $p_{3} =0$. These solutions indicate the state above the Curie temperature ($\theta>\theta_{C}$), when the material exhibits its paraelectric characteristics and the polarization vanishes in the absence of an electric field. The current stage is referred to as the cubic phase, and the formula for this cubic phase may be represented as:
\begin{eqnarray}
Cubic & : & p_{1}=p_{3}=0.\label{A21}
\end{eqnarray}
Below the Curie temperature, the piezoelectric material displays tetragonal, orthorhombic, and rhombohedral phases. In a two-dimensional plane, the tetragonal phase and orthorhombic phase both show polarization in the (0,1) direction. As a result, the solutions for these phases, as given by Equation (\ref{A20}), may be expressed as:
\begin{eqnarray}
Tetragonal & : & p_{1}=0,p_{3}\neq0, \quad \alpha_{1}\frac{\theta_{c}-\theta}{\theta_{c}}+2\alpha_{11}p_{3}^{2}+3\alpha_{111}p_{3}^{4}=0,\label{A22}
\end{eqnarray}
\begin{eqnarray}
Orthorombic & : & p_{1}=0,p_{3}\neq0, \quad \alpha_{1}\frac{\theta_{c}-\theta}{\theta_{c}}+2\alpha_{11}p_{3}^{2}+3\alpha_{111}p_{3}^{4}=0.\label{A23}
\end{eqnarray}
The direction of polarization in rhombohedral phase in $x_{1}-x_{3}$ plane is along the (1,1) direction, therefore the solution for rhombohedral phase is:
\begin{eqnarray}
Rhombohedral & : & p_{1}=p_{3}\neq0, \quad\alpha_{1}+(2\alpha_{11}+\alpha_{12}\frac{\theta_{c}-\theta}{\theta_{c}})p_{3}^{2}+3\alpha_{111}p_{3}^{4}=0.\label{A24}
\end{eqnarray}
The corresponding Landau-Ginzburg-Devonshire energy function for each
phase are as follows: 
\begin{eqnarray}
Cubic & : & W^{2DC}(\theta,\pmb{\overrightarrow{p}})=0,\label{A25}
\end{eqnarray}
\begin{eqnarray}
Tetragonal & : & W^{2DT}(\theta,\pmb{\overrightarrow{p}})=\alpha_{1}\frac{\theta_{c}-\theta}{\theta_{c}}p_{3}^{2}+\alpha_{11}p_{3}^{4}+\alpha_{111}p_{3}^{6},\label{A26}
\end{eqnarray}
\begin{eqnarray}
Orthorombic & : & W^{2DO}(\theta,\pmb{\overrightarrow{p}})=\alpha_{1}\frac{\theta_{c}-\theta}{\theta_{c}}p_{3}^{2}+\alpha_{11}p_{3}^{4}+\alpha_{111}p_{3}^{6},\label{A27}
\end{eqnarray}
\begin{eqnarray}
Rhombohedral & : & W^{2DR}(\theta,\pmb{\overrightarrow{p}})=2\alpha_{1}\frac{\theta_{c}-\theta}{\theta_{c}}p_{3}^{2}+(2\alpha_{11}+\alpha_{12}\frac{\theta_{c}-\theta}{\theta_{c}})p_{3}^{4}+2\alpha_{111}p_{3}^{6}.\label{A28}
\end{eqnarray}

In order to ascertain the values of $\alpha_{1}$, $\alpha_{11}$, $\alpha_{12}$, and $\alpha_{111}$, we examine the experimental data obtained at the Curie point. At this point, the phase shows a cubic phase and zero polarization. Consequently, the Landau-Ginzburg-Devonshire free energy is zero beyond this point, even though $\alpha_{1}$ has a big value. This occurs when $\theta>\theta_{c}$, which is greater than $\theta_{0}$. However, at temperatures somewhat below the Curie temperature, the variable $\alpha_{1}$ becomes negative. By maintaining the values of $\alpha_{11}$, $\alpha_{12}$, and $\alpha_{111}$ at a constant level, the Landau-Ginzburg-Devonshire free energy polynomial function $W^{3D}(\theta,\pmb{\overrightarrow{p}})$ transforms into the equations (\ref{A13}), (\ref{A14}), and (\ref{A15}). Furthermore, at the Curie temperature, there is a balance between the tetragonal and cubic phases, and the expression for $W^{3D}(\theta,\pmb{\overrightarrow{p}})$ may be derived as \citep{Wang2013,Devonshire1949}:

\begin{eqnarray}
W^{3D}(\theta,\pmb{\overrightarrow{p}}) & = &\alpha_{1c}p_{3c}^{2}+\alpha_{11}p_{3c}^{4}+\alpha_{111}p_{3c}^{6}= 0,\label{A29} 
\end{eqnarray}
where the subscript $c$ represents the values at Curie temperature. Now Eq. (\ref{A29}) can be written as:
\begin{eqnarray}
\alpha_{11}p_{3c}^{2}+\alpha_{111}p_{3c}^{4}= -\alpha_{1c},\label{A30} 
\end{eqnarray}
For Curie point rearranging the Eq. (\ref{A9}), we have:
\begin{eqnarray}
2\alpha_{11}p_{3c}^{2}+3\alpha_{111}p_{3c}^{4}= -\alpha_{1c}.\label{A31} 
\end{eqnarray}
Now by solving the Eqs. (\ref{A30}), and (\ref{A31}), the values of $\alpha_{11}$,
$\alpha_{111}$ are calculated as \citep{Wang2013,Devonshire1949}:
\begin{eqnarray}
\alpha_{11} & = & \frac{-2\alpha_{1c}}{p_{3c}^{2}}, \alpha_{111}  =  \frac{\alpha_{1c}}{p_{3c}^{4}}.\label{A32}
\end{eqnarray}
The variable $\alpha_{1}$ exhibits a strong correlation with the reciprocal of the dielectric susceptibility $(\chi)$, and can be determined $\alpha_{1}$
as follows \citep{Wang2013,Devonshire1949}:
\begin{eqnarray}
2\alpha_{1} & = & \frac{1}{\chi}=\frac{(\theta-\theta_{0})}{(\theta_{c}-\theta_{0})}\alpha_{1c}.\label{A33}
\end{eqnarray}
The relationship between $\alpha_{12}$ and $\alpha_{11}$ is expressed as $\alpha_{12}=-a\alpha_{11}$, where $a$ is determined by fitting it to the Landau-Ginzburg-Devonshire free energy term at lower transition temperatures and comparing it to the experimental Landau-Ginzburg-Devonshire free energy at that specific point \citep{Wang2013,Devonshire1949}.

The residual polarization in the absence of an electric field is referred to as spontaneous polarization. This polarization is dependent on temperature and may be described by the following relationship:

The temperature-dependent spontaneous polarization $p_{s}(\theta)$ is obtained by solving Eqs. (\ref{A9}), (\ref{A10}), and (\ref{A11}) with the substitution $p_{3}=p_{s}$ and holding all other variables constant. The expression for $p_{s}(\theta)$ is as follows:

\begin{eqnarray}
Tetragonal & : & p_{s}(\theta) =\sqrt{\frac{\sqrt{\alpha_{11}^{2}-3\alpha_{1}\alpha_{111}\frac{\theta_{c}-\theta}{\theta_{c}}}-\alpha_{11}}{3\alpha_{111}}},\label{A34}
\end{eqnarray}
\begin{eqnarray}
Orthorombic & : &  p_{s}(\theta) =\sqrt{\frac{\sqrt{(2\alpha_{11}+\alpha_{12}\frac{\theta_{c}-\theta}{\theta_{c}})^{2}-12\alpha_{1}\alpha_{111}\frac{\theta_{c}-\theta}{\theta_{c}}}-(2\alpha_{11}+\alpha_{12}\frac{\theta_{c}-\theta}{\theta_{c}})}{6\alpha_{111}}},\label{A35}
\end{eqnarray}
\begin{eqnarray}
Rhombohedral & : & p_{s}(\theta) =\sqrt{\frac{\sqrt{(\alpha_{11}+\alpha_{12}\frac{\theta_{c}-\theta}{\theta_{c}})^{2}-3\alpha_{1}\alpha_{111}\frac{\theta_{c}-\theta}{\theta_{c}}}-(\alpha_{11}+\alpha_{12}\frac{\theta_{c}-\theta}{\theta_{c}})}{3\alpha_{111}}},\label{A36}
\end{eqnarray}

The temperature-dependent spontaneous polarization $p_{s}(\theta)$ is obtained by solving Eqs. (\ref{A22}), (\ref{A23}), and (\ref{A24}) with the substitution $p_{3}=p_{s}$ and holding all other variables constant. The expression for $p_{s}(\theta)$ is as follows:
\begin{eqnarray}
Tetragonal ~and  ~Orthorombic& : & p_{s}(\theta) =\sqrt{\frac{\sqrt{\alpha_{11}^{2}-3\alpha_{1}\alpha_{111}\frac{\theta_{c}-\theta}{\theta_{c}}}-\alpha_{11}}{3\alpha_{111}}},\label{A37}
\end{eqnarray}
\begin{eqnarray}
Rhombohedral & : &  p_{s}(\theta) =\sqrt{\frac{\sqrt{(2\alpha_{11}+\alpha_{12}\frac{\theta_{c}-\theta}{\theta_{c}})^{2}-12\alpha_{1}\alpha_{111}\frac{\theta_{c}-\theta}{\theta_{c}}}-(2\alpha_{11}+\alpha_{12}\frac{\theta_{c}-\theta}{\theta_{c}})}{6\alpha_{111}}},\label{A38}
\end{eqnarray}

The ferroelectric polarization vector $\pmb{\overrightarrow{p}}$ is the order parameter that characterizes the phase transition, as previously stated. The Landau-Ginzburg-Devonshire equation for a vector order parameter $\pmb{\overrightarrow{p}}$ is a semilinear parabolic equation, commonly referred to as a reaction-diffusion system \citep{Kamlah2001,Mueller2008}. It may be defined as follows:
\begin{eqnarray}
\tau\dot{\pmb{\overrightarrow{p}}} & = & \nabla\cdotp(\lambda\nabla\pmb{\overrightarrow{p}})-W^{2D}_{\pmb{\overrightarrow{p}}}(\theta,\pmb{\overrightarrow{p}}),\label{eq:parabolic}
\end{eqnarray}
where $\tau$ and $\lambda$ represent the material characteristics. The first component on the right-hand side of Equation (\ref{eq:parabolic}) represents the uneven distribution of polarization in the nano-scale ferroelectric distribution. This aspect is not taken into account in the current inquiry but might be incorporated in future investigations. To ascertain the consequences of the second law of thermodynamics, it is possible to articulate the dissipation inequality for the Helmholtz free energy function $(\phi)$ by amalgamating the energy balance and entropy inequality.
\begin{eqnarray}
 & (\phi_{\theta}+S)\dot{\theta}+(\phi_{\pmb{\overrightarrow{p}}}-W^{2D}_{\pmb{\overrightarrow{p}}})-\mathcal{D}_{\pmb{\overrightarrow{p}}})\cdotp\dot{\pmb{\overrightarrow{p}}}+(\phi_{\pmb{\varepsilon}}-\pmb{\sigma})\cdotp\dot{\pmb{\varepsilon}}\nonumber \\
 & +(\phi_{\pmb{\nabla\varepsilon}}-\pmb{\hat{\sigma}})\cdotp\nabla\dot{\pmb{\varepsilon}}+(\phi_{\pmb{\overrightarrow{E}}}+\pmb{\overrightarrow{D}})\cdotp\dot{\pmb{\overrightarrow{E}}}+\frac{\pmb{\overrightarrow{q}}}{\theta}\cdotp\nabla\theta\leq0.\label{eq:inequality}
\end{eqnarray}
The derivative of the dissipation potential, denoted as $\mathcal{D}_{\pmb{\overrightarrow{p}}}$, is defined as $\mathcal{D}_{\dot{\pmb{\overrightarrow{p}}}}=\tau\dot{\pmb{\overrightarrow{p}}}$, where $\tau$ represents a parameter. This definition is enough to fulfill the dissipation inequality stated in Equation (\ref{eq:Rest_eq}). The symbol $\pmb{\overrightarrow{q}}$ represents the heat flux vector, and $S$ represents the entropy. In addition, the subsequent constitutive equations may be obtained from Equation (\ref{eq:free_energy}):
\begin{eqnarray}
\phi_{\pmb{\varepsilon}}=\pmb{\sigma}=\pmb{C}(\pmb{\varepsilon}(\pmb{\overrightarrow{u}})-\pmb{\varepsilon^{t}}(\pmb{\overrightarrow{p}}))-\pmb{e}\pmb{\overrightarrow{E}}-\pmb{\beta}(\theta-\theta_{R}),\label{eq:Stress}
\end{eqnarray}
\begin{eqnarray}
\phi_{\pmb{\nabla\varepsilon}} & = & \pmb{\hat{\sigma}}=\pmb{\mu}\pmb{\overrightarrow{E}},\label{eq:flexo}
\end{eqnarray}
\begin{eqnarray}
-\phi_{\pmb{\overrightarrow{E}}} =  \pmb{\overrightarrow{D}}&=&\pmb{\epsilon}\pmb{\overrightarrow{E}}+\pmb{e}(\pmb{\varepsilon}(\pmb{\overrightarrow{u}})-\pmb{\varepsilon^{t}}(\pmb{\overrightarrow{p}}))+\pmb{\eta}(\theta-\theta_{R})\nonumber \\
&+&\pmb{\mu}(\pmb{\nabla\varepsilon}(\pmb{\overrightarrow{u}})-\pmb{\nabla\varepsilon^{t}}(\pmb{\overrightarrow{p}}))+\pmb{\overrightarrow{p}},\label{eq:ED}
\end{eqnarray}
\begin{eqnarray}
-\phi_{\theta} & = & S=\pmb{\beta}\pmb{\varepsilon}(\pmb{\overrightarrow{u}})-\pmb{\eta}\pmb{\overrightarrow{E}}+W_{\theta}(\theta,\pmb{\overrightarrow{p}}),\label{eq:entropy}
\end{eqnarray}
\begin{eqnarray}
 &  & \phi_{\pmb{\overrightarrow{p}}}=W^{2D}_{\theta,\pmb{\overrightarrow{p}}}(\pmb{\overrightarrow{p}}),\pmb{\overrightarrow{q}}=-\pmb{k}\nabla\theta,\pmb{\overrightarrow{E}}=-\nabla V,\mathcal{D}_{\dot{\pmb{\overrightarrow{p}}}}\cdotp\dot{\pmb{\overrightarrow{p}}}\geq0,\pmb{k}>0.\label{eq:Rest_eq}
\end{eqnarray}
The condition $\mathcal{D}_{\dot{\pmb{\overrightarrow{p}}}}\cdotp\dot{\pmb{\overrightarrow{p}}}\geq0$ implies that $\tau$ is greater than zero. Now, we modify the set of balancing equations by employing the aforementioned constitutive rules. The governing equations for the 2D composite system on the $x_{1}-x_{3}$ plane, which describe the model, can be defined as follows.
\begin{eqnarray}
\rho\ddot{\pmb{\overrightarrow{u}}} & = &  \nabla\cdotp(\pmb{\sigma}-\pmb{\hat{\sigma}})+\pmb{\overrightarrow{F}},\label{eq:Mechanical_ge}
\end{eqnarray}
\begin{eqnarray}
\tau\dot{\pmb{\overrightarrow{p}}} & = & -\phi_{\pmb{\overrightarrow{p}}},\label{eq:parabolic-1}
\end{eqnarray}
\begin{eqnarray}
-\theta\dot{{\phi_{\theta}}} & = & \tau\dot{\pmb{\overrightarrow{p}}}^{2}+r+\nabla\cdotp(\pmb{k}\nabla\theta),\label{eq:heat}
\end{eqnarray}
\begin{eqnarray}
\nabla\cdotp \pmb{\overrightarrow{D}} & = & \rho_{e},\label{eq:Electric_ge}
\end{eqnarray}
The variables in the equation are defined as follows: $\rho$ represents the mass density, $\rho_{e}$ represents the electric charge density, $r$ represents the external heat supply to the system, $\pmb{k}$ represents thermal conductivity, and $\pmb{\overrightarrow{F}}$ represents the external body force term. Furthermore, the fluctuations in the inertia and body forces at the micro-scale, in relation to their average values throughout the Representative Volume Element (RVE), might potentially impact the equilibrium problem at the micro-scale and the resulting homogenized stress. When considering mechanical significance, the volume average is only significant on a large scale. The microstructure is much less than the frequency of the external load, as a quasistatic load is being applied in this case. In addition, the process of phase change does not include micro-inertia \citep{deSouzaNeto2015}. Hence, the research disregards the impact of micro-inertia terms and body forces. Consequently, Equation \ref{eq:Mechanical_ge} may be expressed as:

\begin{eqnarray}
 \nabla\cdotp(\pmb{\sigma}-\pmb{\hat{\sigma}})& = &  0.\label{eq:Mechanical_ge2d}
\end{eqnarray}

\subsection{Mesoscale phase-field thermo-electromechanical model}

In this section, an alternative phenomenological explanation for the previously mentioned piezo-ferroelectric material is introduced. Instead of utilizing Equation (\ref{eq:parabolic}), we substitute it with a scalar Ginzburg-Landau equation that characterizes the amplitude of the spontaneous polarization.
Furthermore, we present a mathematical equation that describes the development of the unit polarization vector, represented as $\pmb{\overrightarrow{n}}$ \citep{Zhang2005}:

\begin{eqnarray}
\pmb{\overrightarrow{p}} & = & \left|\pmb{\overrightarrow{p}}\right|\pmb{\overrightarrow{n}},\label{eq:macro-Pol}
\end{eqnarray}
The proposed isotropic Ginzburg-Landau potential, denoted as $W(\theta,\pmb{\overrightarrow{p}})$, which contradicts the anisotropic function described in Equation (\ref{eq:polynomial}) \citep{Wang2004}.

\begin{eqnarray}
W^{3D}(\theta,\pmb{\overrightarrow{p}})  & = & \alpha_{1}\frac{\theta_{c}-\theta}{\theta_{c}}p^{2}+\alpha_{11}p^{4},\label{eq:pol2-1}
\end{eqnarray}
here, the polarization vector $p_{1}=p_{2}=p_{3}=p$ is determined by the isotropic assumption and the Landau-Ginzberg-Devonshire function. It becomes a function of the scalar variable $p$, which represents the magnitude of the polarization vector $\pmb{\overrightarrow{p}}$. In addition, we propose a free energy function that is identical to the previous one, except for the inclusion of the scalar gradient $p$ (see to Equation (\ref{eq:free_energy})).

\begin{eqnarray}
\phi(\pmb{\overrightarrow{E}},\pmb{\varepsilon}(\pmb{\overrightarrow{u}}),p,\nabla p,\theta) & = & \frac{1}{2}\pmb{\lambda}\left|\nabla p\right|^{2}+W^{3D}(\theta,\pmb{\overrightarrow{p}})+\frac{1}{2}\pmb{C}\pmb{\varepsilon}^{el}:\pmb{\varepsilon}^{el}+ (\theta-\theta_{R})\pmb{\beta}\pmb{\varepsilon}-\frac{1}{2}\pmb{\overrightarrow{E}}^{T}\pmb{\epsilon}\pmb{\overrightarrow{E}}-\pmb{\mu}\pmb{\overrightarrow{E}}\nabla\pmb{\varepsilon}^{el}\nonumber \\
 & - & \pmb{e}\pmb{\overrightarrow{E}}\pmb{\varepsilon}^{el}-p\pmb{\overrightarrow{E}}-\pmb{\eta}(\theta-\theta_{R})\pmb{\overrightarrow{E}}.\label{eq:phase-GE-1}
\end{eqnarray}

The symbol $\pmb{\overrightarrow{E}}$ represents the vector for electric field intensity, whereas $\pmb{\varepsilon}(\pmb{\overrightarrow{u}})$ represents the vector function for electric field intensity.The symbol $\pmb{\varepsilon}(\pmb{\overrightarrow{u}})$ indicates the overall mechanical strain, whereas $p$ represents the scalar value of spontaneous polarization. $\theta$ represents the temperature, and $\theta_{R}$ represents the reference temperature.From a modelling standpoint, adopting an isotropic Ginzburg-Landau potential necessitates employing a more macroscopic methodology, as the anisotropy directions of the potential (Eq. (\ref{eq:polynomial})) coincide with the crystal orientations. All the constitutive relations remain unchanged, except for the substitution of the polarization vector $\pmb{\overrightarrow{p}}$ with the scalar $p$ in Eqs.(\ref{eq:Stress}), (\ref{eq:ED}), (\ref{eq:entropy}), and (\ref{eq:Rest_eq}). The modified equations are as follows:

\noindent 
\begin{eqnarray}
\phi_{\pmb{\varepsilon}}=\pmb{\sigma}=\pmb{C}(\pmb{\varepsilon}(\pmb{\overrightarrow{u}})-\pmb{\varepsilon}^{t}(p))-\pmb{e}\pmb{\overrightarrow{E}}-\pmb{\beta}(\theta-\theta_{R}),\label{eq:Stress-1}
\end{eqnarray}
\noindent 
\begin{eqnarray}
\phi_{\nabla\pmb{\varepsilon}} & = & \pmb{\hat{\sigma}}=\pmb{\mu}\pmb{\overrightarrow{E}}\label{eq:flexo-1}
\end{eqnarray}
\noindent 
\begin{eqnarray}
-\phi_{\pmb{\overrightarrow{E}}}=\pmb{\overrightarrow{D}}=\pmb{\epsilon}\pmb{\overrightarrow{E}}+\pmb{e}(\pmb{\varepsilon}(\pmb{\overrightarrow{u}})-\pmb{\varepsilon}^{t}(p))+\pmb{\eta}(\theta-\theta_{R})
-\pmb{\mu}(\pmb{\varepsilon}(\pmb{\overrightarrow{u}})-\pmb{\varepsilon}^{t}(p))+p\label{eq:ED-1}
\end{eqnarray}
\noindent 
\begin{eqnarray}
-\phi_{\theta} & = & S=\pmb{\beta}(\pmb{\varepsilon}(\pmb{\overrightarrow{u}}))-\pmb{\eta}\pmb{\overrightarrow{E}}-a_{\theta}(\theta-\theta_{R})+W_{\theta}(\theta,p),\label{eq:entropy-1}
\end{eqnarray}
\noindent 
\begin{eqnarray}
\phi_{\nabla p} & = & \pmb{\lambda}\nabla p,\pmb{\overrightarrow{Q}}=-\pmb{k}\nabla\theta,k>0,\label{eq:Rest_eq-1}
\end{eqnarray}
and governing equations for Mechanical and electrical field remains
the same as given by Eqs. (\ref{eq:Mechanical_ge}), and (\ref{eq:Electric_ge})
whereas, the governing equations for polarization field and thermal
field changes to:
\noindent 
\begin{eqnarray}
\tau\dot{p} & = & \nabla\cdotp(\pmb{\lambda}\nabla p)-\phi_{p},\label{eq:parabolic-1-1}
\end{eqnarray}
\noindent 
\begin{eqnarray}
-\theta\dot{\overline{\phi_{\theta}}} & = & \pmb{\tau}\dot{p}^{2}+r+\nabla\cdotp(\pmb{k}\nabla\theta)\label{eq:heat-1}
\end{eqnarray}

\subsection{Viscoelastic damped thermo-electromechanical model}

The viscoelasticity is also a property exhibited by certain lead-free
piezoelectric materials due to which the material restores some of
strain for next cycle of loading. Therefore, the stress-strain relationship
is no more linear for material and a temporal strain has been stored
in the material which cause the gibbs energy function to be modified
accordingly. The equation for free energy ($\phi$) considering three
independent variables $\pmb{\chi},$ $\pmb{\overrightarrow{E}},$ and $\theta$ of our system has
been of the following form \citep{Chandrasekharaiah1988}:
\begin{eqnarray}
\rho\phi(\pmb{\chi},\pmb{\overrightarrow{E}},\theta) & = & \frac{1}{2}\pmb{C}\pmb{\chi}:\pmb{\chi}-\pmb{e}\pmb{\overrightarrow{E}}\pmb{\chi}-\frac{1}{2}\pmb{\epsilon}\pmb{\overrightarrow{E}}^{2}-\pmb{\mu}\pmb{\overrightarrow{E}}\nabla\pmb{\chi}+\pmb{\overrightarrow{p}}\pmb{\overrightarrow{E}}-\pmb{\beta}\theta\pmb{\chi}-\pmb{\eta}\pmb{\overrightarrow{E}}\theta-\frac{\rho}{2}a_{\theta}\theta^{2},\label{eq:Visc-GE}
\end{eqnarray}
Where, $\pmb{C},$ $\pmb{\epsilon}$, $\pmb{e},$ $\pmb{\mu}$, $\pmb{\beta}$,
$\pmb{\eta}$, and $a_{\theta}$ are the elastic permittivity, dielectric,
piezoelectric, flexoelectric, thermomechanical, thermoelectric, heat
capacity coefficients respectively and $a_{\theta}=C_{\varepsilon}^{V}/\theta_{R}$
where $C_{\varepsilon}^{V}$ is the heat capacity at constant volume
and elastic strain.
The constitutive equations for thermo-electro-mechanical coupling
are as follows \citep{Patil2009a}:
\begin{eqnarray}
\pmb{\sigma} & = & \phi_{\pmb{\chi}}=\pmb{C}\pmb{\chi}-\pmb{e}\pmb{\overrightarrow{E}}-\pmb{\beta}\theta,\label{eq:Vis-sigma}
\end{eqnarray}
\begin{eqnarray}
\pmb{\hat{\sigma}} & = & \phi_{\nabla\pmb{\chi}}=\pmb{\mu}\pmb{\overrightarrow{E}},\label{eq:Vis-Flexo}
\end{eqnarray}
\begin{eqnarray}
D_{i} & = & -\phi_{\pmb{\overrightarrow{E}}}=\pmb{\epsilon}\pmb{\overrightarrow{E}}+\pmb{e}\pmb{\chi}+\pmb{\eta}\theta-\pmb{\mu}\nabla\pmb{\chi}+\pmb{\overrightarrow{p}},\label{eq:Visco-Ed}
\end{eqnarray}
\begin{eqnarray}
\rho S & = & - \phi_{\theta}=\pmb{\beta}\pmb{\chi}+\pmb{\eta}\pmb{\overrightarrow{E}}+\rho a_{\theta}\theta.\label{eq:Visc-Ent}
\end{eqnarray}
Where, $S$ is the entropy, $\pmb{\chi}$ is the memory dependent strain,
and $\pmb{\overrightarrow{p}}$ is the spontaneous polarization with in the material
which is used for switching of ferroelectric domains during phase
change and dependent on the temperature. The memory dependent strain
$\pmb{\chi}$ is adopted to replace the elastic strain to display viscoelastic
behavior arising from microscale phenomenon of hysteresis, memory
and fractional order dynamics and so on. The strain field $\pmb{\chi}$
is defined as \citep{Youssef2016,Li2020a}:
\begin{eqnarray}
\pmb{\chi} & = & (1+\tau_{\varepsilon}^{\gamma_{\varepsilon}}D_{t}^{\gamma_{\varepsilon}})\pmb{\varepsilon},(0<\gamma_{\varepsilon}\le1),\label{eq:Visco-strain}
\end{eqnarray}
Where, $\tau_{\varepsilon}$ is the strain relaxation time parameter
and $\gamma_{\varepsilon}$ is the fractional order parameter of elastic
strain. In Eq. (6), $\pmb{\varepsilon}$ is the elastic strain and
$\tau_{\varepsilon}{\gamma_{\varepsilon}}D_{t}^{\gamma_{\varepsilon}}\pmb{\varepsilon}$
is the fractional order strain can be applied to accurately characterize
material hysteresis and memory dependence feature of elastic deformation
\citep{DiPaola2011}. The definition of Caputo time-fractional derivative
is given as below \citep{Magin2010,Xiao2016}:
\begin{eqnarray}
D_{t}^{\gamma_{\varepsilon}}f(t) & = & \frac{1}{\Gamma(1-\gamma_{\varepsilon})}\int_{0}^{t}\frac{\partial f}{\partial t}\frac{1}{(t-\tau)^{\gamma_{\varepsilon}}}d\tau,(0\le\gamma_{\varepsilon}\le1),\label{eq:temporal}
\end{eqnarray}
The spontaneous polarization $\pmb{\overrightarrow{p}}$ is defined as:
\begin{eqnarray}
\pmb{\overrightarrow{p}} & = & \epsilon_{0}\chi_{e}\pmb{\overrightarrow{E}}+\epsilon_{0}\chi_{e}^{2}\pmb{\overrightarrow{E}}^{2}+\epsilon_{0}\chi_{e}^{3}\pmb{\overrightarrow{E}}^{3}+\cdots\ldots\ldots\ldots.\label{eq:polarization}
\end{eqnarray}
Where, $\chi_{e}$ is the dielectric susceptibility which is defined
as $\chi_{e}=\epsilon_{r}-1$, where $\epsilon_{r}$ is the relative
dielectric permittivity. Dielectric susceptibility is temperature
dependent and its relation with temperature can be written as \citep{Trainer2000}:
\begin{eqnarray}
\chi_{e} & = & \frac{3\theta_{c}}{\theta-\theta_{c}},\label{eq:suscept}
\end{eqnarray}
where $\theta_{c}$ is the Curie temperature. However, higher order terms
in Eq.(8) may be neglected on the basis of contribution and Eq.(8)
can be reduced as:
\begin{eqnarray}
\pmb{\overrightarrow{p}} & = & \epsilon_{0}\chi_{e}\pmb{\overrightarrow{E}}.\label{eq:pol2}
\end{eqnarray}
Gradient equations correspond to the relationships between the linear
strain and mechanical displacement, the electric field and electric
potential, and the thermal field \citep{Sagi2012,Narayan2002,Calvo2007}
and temperature change. They are stated respectively as:
\begin{eqnarray}
\pmb{\varepsilon} & = & \frac{1}{2}(\nabla\pmb{\overrightarrow{u}}+\nabla\pmb{\overrightarrow{u}}^{T}),\label{eq:strain}
\end{eqnarray}
\begin{eqnarray}
\pmb{\overrightarrow{Q}} & = & -\pmb{k}\nabla\theta-\frac{\tau_{Q}^{\alpha_{Q}}}{\alpha_{Q}!}\frac{\partial^{\alpha_{Q}}}{\partial t^{\alpha_{Q}}}\pmb{\overrightarrow{Q}},\label{eq:heatflux}
\end{eqnarray}
where $\pmb{\varepsilon}$, $\pmb{\overrightarrow{E}},$ $\pmb{\overrightarrow{Q}},$ $\pmb{\overrightarrow{u}},$ $V$ and $\theta$
are the strain tensor, electric field vector, thermal
field vector, mechanical displacement vector, electric potential and
temperature change from the reference, respectively. Moreover, $\tau_{Q}$
is the thermal relaxation time, \pmb{k} is thermal conductivity, $\alpha_{Q}$
is fractional order parameter of the heat transfer and defined as:
\begin{eqnarray}
\frac{\partial^{\alpha_{Q}}}{\partial t^{\alpha_{Q}}}f(x,t) & = & \begin{cases}
f(x,t)-f(x,0),~\mbox{when}~  \alpha_{Q}\rightarrow0,\\
I^{\alpha_{Q}-1)}\frac{\partial f(x,t)}{\partial t}, ~\mbox{when}~    0<\alpha_{Q}<1,\\
\frac{f(x,t)}{\partial t}, ~\mbox{when}~    \alpha_{Q}=1.
\end{cases}\label{eq:fractional-op}
\end{eqnarray}
Where Riemann-Liouville fractional integral operator $I^{\alpha_{Q}}$
is commonly defined as:
\begin{eqnarray}
(I^{\alpha_{Q}})f(t) & = & \frac{1}{\Gamma(\alpha_{Q})}\int_{0}^{t}(t-\tau)^{\alpha-1}d\tau.\label{eq:Riemann}
\end{eqnarray}
The linear fundamental equations for the thermo-electromechanical
structure occupying volume, under steady-state conditions, can be
summarized as follows:
\begin{eqnarray}
\nabla\cdot(\pmb{\sigma}-\pmb{\hat{\sigma}})+\rho \pmb{\overrightarrow{F}} & = & \rho\ddot{\pmb{\overrightarrow{u}}},\label{eq:GE-visco}
\end{eqnarray}
\begin{eqnarray}
\nabla\cdot\pmb{\overrightarrow{D}} & = & 0,\label{eq:ge-ed}
\end{eqnarray}
\begin{eqnarray}
-\nabla\cdot\pmb{\overrightarrow{Q}}+\rho Q_{gen} & = & \rho \theta_{R}\dot{S},\label{eq:ge-temp}
\end{eqnarray}
where $\pmb{\overrightarrow{F}}$ is the body force term which is assumed to be vanished
in our model. Substituting all constitutive and complementary equations,
the governing equation can be reduced to as following:
\begin{eqnarray}
\pmb{C}(1+\tau_{\varepsilon}^{\gamma_{\varepsilon}})D_{t}^{\gamma_{\varepsilon}})\nabla\pmb{\varepsilon}-\pmb{e}\nabla \pmb{\overrightarrow{E}}-\pmb{\beta}\nabla\theta
-\pmb{\mu}\nabla\pmb{\overrightarrow{E}}+\rho \pmb{\overrightarrow{F}} & = & \rho\ddot{\pmb{\overrightarrow{u}}}\label{eq:ME}
\end{eqnarray}
\begin{eqnarray}
\pmb{\epsilon}\nabla\pmb{\overrightarrow{E}}+\pmb{e}(1+\tau_{\varepsilon}^{\gamma_{\varepsilon}}D_{t}^{\gamma_{\varepsilon}})\nabla\pmb{\varepsilon}+\pmb{\eta}\nabla\theta
-\pmb{\mu}(1+\tau_{\varepsilon}^{\gamma_{\varepsilon}}D_{t}^{\gamma_{\varepsilon}})\nabla^{2}\pmb{\varepsilon}+\nabla\pmb{\overrightarrow{p}} & = & 0\label{eq:EE}
\end{eqnarray}
\begin{eqnarray}
(1+\frac{\tau_{Q}^{\alpha_{Q}}}{\alpha_{Q}!}\frac{\partial^{\alpha_{Q}}}{\partial t^{\alpha_{Q}}})[\theta_{R}\pmb{\beta}(1+\tau_{\varepsilon}^{\gamma_{\varepsilon})}D_{t}^{\gamma_{\varepsilon}})\dot{\pmb{\varepsilon}}
+\theta_{R}\pmb{\eta}\dot{\pmb{\overrightarrow{E}}}+\rho C_{\varepsilon}^{V}\theta-\rho Q_{gen.}] & = & -\pmb{k}\nabla^{2}\theta\label{eq:Visc-HE}
\end{eqnarray}
\subsection{Effective properties estimation}

The development of high-performance piezoelectric materials requires a robust theoretical framework and advanced numerical tools to estimate effective properties based on the underlying microstructure. This challenge, commonly referred to as the homogenization problem, is well-documented in the literature (e.g., \cite{MiltonBook,torquato2002}) and involves deriving effective material coefficients that encapsulate the complex interactions within heterogeneous microstructures. The homogenization framework not only enables the solution of the governing PDEs with homogenized coefficients but also facilitates predictive modeling of a microstructured piezoelectric material’s macroscopic response under controlled microstructural modifications. This approach is essential for material design, providing insights into how specific microstructural parameters—such as phase distribution, orientation, and connectivity—impact macroscopic performance metrics.

This methodology relies fundamentally on the Hill–Mandel Principle of Macrohomogeneity and the assumption of scale separation, which allows decoupling of the differential equations at the macro- and microscales. The principle ensures that the energy at the microscale accurately represents the macroscopic behavior, while scale separation assumes that the microscale structure is significantly smaller than the macroscopic dimensions, justifying the averaging process.

Intuitively, the estimation of effective properties can be formulated as a relationship between the averages of the microscopic fields, with each field satisfying its respective differential constraints. For instance, in the electromechanical problem presented in Section 4.1, the microfields are subject to the differential restrictions given by equations (\ref{eq:Eq6})-(\ref{eq:Eq10}). These can be computed numerically for a specified microstructured sample. Under appropriate boundary conditions, the corresponding volume averaged fields can be determined, allowing for the estimation of effective coefficients. In this example, the coefficients $\boldsymbol{C}, \boldsymbol{\epsilon}, \boldsymbol{e}, \boldsymbol{G}$, and $\boldsymbol{\mu}$ appear in Eq. (\ref{eq:EMC}), and correspond to the elastic, permittivity, piezoelectric, electrostrictive, and flexoelectric properties, respectively. Note that nonlinear evolution of the properties can also be captured in this framework, from which linear coefficients are obtained after lineralization. 

For the problem presented in Section 4.1, we can determine the composite's effective characteristics by applying uniform gradient boundary conditions over a microstructural sample with a sufficiently representative amount of heterogeneities. The Voigt notation is used to convert four- and three-digit indices of elastic and piezoelectric coefficients into a two-index notation, simplifying the notation system and reducing the number of variables to streamline the computational process. The volume average of a quantity $A$ is denoted as $ \langle A \rangle $ in the following calculation, and it can be computed numerically as follows \citep{MiltonBook,torquato2002}:
\begin{equation}
\langle A \rangle = \frac{1}{|\Omega|} \int_{\Omega} A \, \mathrm{d}V,
\end{equation}
where \( |\Omega| \) denotes the total volume of the sample domain \( \Omega \) across which the integration is performed. By imposing suitable boundary conditions for displacements and electric potential, such that the entire microstructural sample is subjected to uniform infinitesimal strains and constant electric fields, effective elastic stiffness at constant electric field, and piezoelectric strain coupling coefficients can be obtained.

In a two-dimensional setting for the plane $x_1-x_3$, under uniform field gradient boundary conditions given by: 
\begin{eqnarray}
\left\langle \varepsilon_{11} \right\rangle & = & \bar{\varepsilon}_{11},~\left\langle \varepsilon_{33} \right\rangle=0,~\left\langle \varepsilon_{13} \right\rangle=0,~\left\langle E_{33} \right\rangle=0,\label{eq:Eq13}
\end{eqnarray}
where $\bar{\varepsilon}_{11}$ is a constant applied strain, the composite's effective coefficients are derived as the following
relations between volume averages \citep{Krishnaswamy2020,Saputra2017}:
\begin{equation}
e_{31, e f f .}=\frac{\left\langle D_3\right\rangle}{\bar{\varepsilon}_{11}}, \quad c_{11, e f f .}=\frac{\left\langle\sigma_{11}\right\rangle}{\bar{\varepsilon}_{11}}, \quad c_{13, e f f .}=\frac{\left\langle\sigma_{33}\right\rangle}{\bar{\varepsilon}_{11}},
\end{equation}
where $\left\langle D_{3}\right\rangle $ is the volume average of
the electric flux density component $D_{3}$, and $\left\langle \sigma_{11}\right\rangle$ and $\left\langle \sigma_{33}\right\rangle$ are the volume-averaged stress components in the respective directions.

Similarly, under the conditions
\begin{eqnarray}
\left\langle \varepsilon_{11} \right\rangle & = & 0,~\left\langle \varepsilon_{33} \right\rangle=\bar{\varepsilon}_{33},~\left\langle \varepsilon_{13} \right\rangle=0,~\left\langle E_{33} \right\rangle=0,\label{eq:Eq15}
\end{eqnarray}
the composite's effective coefficients are derived as \citep{Krishnaswamy2020,Saputra2017}:
\begin{equation}
e_{33, e f f .}=\frac{\left\langle D_3\right\rangle}{\bar{\varepsilon}_{33}}, \quad c_{33, e f f .}=\frac{\left\langle\sigma_{33}\right\rangle}{\bar{\varepsilon}_{33}}, \quad c_{13, e f f .}=\frac{\left\langle\sigma_{11}\right\rangle}{\bar{\varepsilon}_{33}}.
\end{equation}
When the microstructure is geometrically periodic—that is, it can be constructed by tessellating a given unit cell-periodic boundary conditions are typically assumed. In contrast, if the microstructure is random, a Representative Volume Element (RVE) analysis is required with appropriate boundary conditions. However, even in this case, periodic boundary conditions have been shown to accelerate the convergence of the RVE analysis, allowing for more realistic or smaller randomly microstructured samples and, consequently, reducing computational cost \cite{INDERGAND2023105426}. Numerical methods based on fast Fourier transforms are particularly well-suited for periodic boundary conditions \cite{MOULINEC199869}. In the following sections, we summarize the main numerical techniques that can be used to solve the homogenization problem for a given microstructured sample.

\subsection{Solution methodologies}

All of the governing equations in the above discussed models are partial differential equations (PDEs) and numerical methods to solve these PDEs are required because the analytical solutions are available for only specific cases. Hence, some common numerical techniques to solve PDEs are given below:

\subsubsection{Finite difference method (FDM)}

Finite difference methods transform nonlinear ordinary differential equations (ODEs) or partial differential equations (PDEs) into a linear equational system amenable to solution using matrix algebra approaches \cite{LeVeque2007}. Contemporary computers are capable of efficiently doing linear algebra calculations, which, along with their rather simple implementation, has resulted in the extensive usage of Finite Difference Methods (FDMs) in current numerical analysis \cite{Strang2007}. Contemporary FDMs are widely used for numerically solving PDEs, alongside finite element methods \cite{Morton2005}. In numerical treatment of a partial differential equation using a FDM, the differentiable solution is approximated by a grid function. This function is defined only at a finite number of grid points located in the underlying domain and its boundary. Every derivative included in the partial differential equation must be substituted with an appropriate divided difference of function values at the selected grid locations \cite{Gerald2004}.

Approximations of derivatives using difference formulae can be derived by several approaches, such as a Taylor expansion, local balancing equations, or a suitable interpretation of finite difference methods as tailored finite element methods \cite{Thomas2004}. In the literature, the first two of these three approaches are commonly referred to as finite difference methods (in their original semantic sense) and finite volume methods, respectively \cite{Trefethen2000}.

Within the finite differences approach, one adheres to the following principles: 
\begin{itemize}
    \item The domain of the given differential equation must include a substantial number of test points (grid points);
    \item All derivatives needed at grid points will be substituted with approximations of finite differences that utilise values of the grid function at adjacent grid points.
\end{itemize}

Solutions to partial differential equations must satisfy boundary and/or beginning conditions. Different from start and boundary value issues in ordinary differential equations, the geometry of the underlying domain is now a significant factor.

FDM accuracy is influenced by the grid spacing size, $h$. Reducing the value of $h$ often enhances precision, but it also results in a higher quantity of grid points, therefore substantially increasing the computational expense \cite{Zienkiewicz2005}. FDM also introduces the high truncation error as it approximates derivatives as finite differences. Due to the assumption of uniform grids, FDM struggles with complex or aberrant geometries. While FDM can handle basic Dirichlet or Neumann boundary conditions, more complex ones like mixed or time-varying boundary conditions might be difficult to implement, leading to errors or inefficiencies \cite{Strikwerda2004}. For higher-dimensional problems like 2D or 3D PDEs, FDM need solving large systems of equations simultaneously. This can be memory- and computational-intensive for fine grids. Efficient techniques like sparse matrix approaches are needed to solve these complicated problems. Even these approaches may be slow or computationally intensive. Iterative approaches like Newton's method are often used to solve nonlinear PDEs utilizing FDM. Every cycle resolves a large system of linear equations, increasing computational complexity \cite{Hirsch2007}. Adaptive mesh refinement (AMR) improves grid resolution on steep slopes or complicated features. FDM is less flexible than FEM or FVM in managing adaptive grids, limiting its ability to handle localized feature concerns \cite{Thomas2004}. In solutions with significant spatial changes, FDM may have trouble resolving boundary layers. Adaptive mesh refinement or higher-order techniques are needed to capture these sudden fluctuations, but traditional FDM does not contain them. Errors at any grid point might propagate over the solution, especially in time-dependent applications. Minor inaccuracies can add up to large deviations from the solution \cite{Morton2005}.

Despite these constraints, the finite difference method continues to be a valuable and extensively used technique in solving various forms of differential equations because of its simplicity and straightforward implementation in regular domains.

 \subsubsection{Finite volume method (FVM)}

 The finite volume method (FVM) is a numerical approach employed for solving PDEs, especially those exhibiting conservation laws such as fluid flow, heat transfer, and mass transport. FVM is extensively employed in computational fluid dynamics (CFD) due to its ability to guarantee the preservation of variables like mass, momentum, or energy across the control volumes in the discretized domain \cite{versteeg2007introduction}.

The domain is partitioned into compact, non-redundant control volumes (CVs). In contrast to the finite difference technique, which evaluates the solution at discrete points, or the finite element approach, which evaluates it within elements, the FVM solves the governing equations across control volumes. Formulation of FVM is founded upon the fundamental concept of flux conservation. The flux entering a control volume must be equal to the sum of the flux exiting the control volume and any source or sink terms enclosed inside the volume. Thus, the conservation of the quantity being modelled (such as mass, momentum, or energy) is guaranteed. The formulation of FVM begins with the integral form of the governing PDEs, which is obtained immediately from the conservation rules. The equation is resolved by integrating its differential form across each control volume \citep{Hirsch2007}.

An extensively employed numerical method for solving PDEs, notably in the fields of fluid dynamics and heat transfer, is the FVM. Although FVM has certain benefits, such as its conservation features and suitability for intricate geometries, it also has some constraints. These constraints stem from its formulation, numerical execution, and obstacles particular to the topic. Analysis of complexity in unstructured meshes, numerical diffusion (artificial diffusion) in situations dominated by advection, constrained to lesser levels of precision without intricate reconstructions, challenges in managing intricate boundary circumstances, destabilization problems with explicit time-stepping under the CFL condition, solutions to higher-order partial differential equations, coupling in multiphysics: an analysis of complexity, evaluation of mesh quality and resolution sensitivity, complex geometry in 3D situations result in high processing costs, and inadequacy for non-conservative equation modelling are some of its limitations \cite{moukalled2016finite}.

Notwithstanding these constraints, the FVM method is extensively employed because of its strong handling of conservation principles, its adaptability in terms of geometry, and its capability to simulate intricate physical events. Nevertheless, in some scenarios (particularly those that need great precision, handle intricate boundaries, or include equations of higher order), other approaches such as FEM or spectral techniques may be more suitable \cite{patankar1980numerical, fletcher1991computational, ferziger2002computational, wesseling2001introduction, reddy2006introduction}.

\subsubsection{Finite element method (FEM)}

The finite element method (FEM) is a robust numerical approach employed for the purpose of obtaining approximate solutions to intricate boundary value problems that encompass PDEs or integral equations \cite{moaveni2015finite}. Due to its adaptability in dealing with complicated geometries and boundary conditions, FEM is extensively utilized in disciplines such as structural mechanics, fluid dynamics, heat transport, and electromagnetics \cite{karniadakis2005spectral, zienkiewicz2005finite}.

The FEM is a robust and adaptable numerical method extensively employed for solving PDEs, particularly in the fields of structural analysis, heat transfer, fluid dynamics, and electromagnetics \cite{hughes2012finite}. Nevertheless, despite its numerous benefits, FEM has significant constraints and restrictions that could impact its effectiveness, precision, or suitability in specific circumstances.

Mesh creation may be challenging, especially when dealing with intricate geometries and 3D difficulties \cite{bathe1996finite}. Some disadvantages of this system include high memory and processing power demands for large-scale or high-resolution issues, convergence of nonlinear problems, need for iterative solutions, instability of time-stepping approaches, and considerable computational cost \cite{korn2000mathematical}. Moreover, coupling various physical phenomena might be difficult and demanding in terms of resources. Fails to accurately capture abrupt breaks without specialist techniques. Inadequate numerical integration, especially for higher-order components, can give rise to errors \cite{quarteroni2008numerical}. Application of intricate or non-uniform boundary conditions poses challenges. At the element level, FEM is not intrinsically conservative. Inadequate management of very thin structures or layers lacking certain components. Considerable effort is necessary for the understanding and display of data.

Even when paired with adaptive meshing, nonlinear solvers, and improved element formulations, FEM continues to be one of the most extensively utilized and flexible approaches for handling a wide range of engineering and scientific issues. However, for specific categories of problems, other approaches such as FVM, Boundary Element Method (BEM), or Spectral Methods may be more appropriate \cite{versteeg2007introduction}.

\subsubsection{Spectral solution scheme}

Spectral or Fast Fourier Transform (FFT)-based homogenization approaches have evolved significantly, particularly in addressing the complexities of electromechanical coupling in ferroelectric materials. Originating from the fixed-point iteration scheme proposed by Moulinec and Suquet \cite{MOULINEC199869} for purely mechanical problems, these methods have been extended to accommodate the multifaceted interactions inherent in ferroelectrics. The foundation of these approaches lies in the strong formulation of equilibrium, leveraging multifield Green’s functions \cite{BuroniSaez} of a reference medium computed in the transformed (Fourier) space. The governing problem is typically cast as a Lippmann–Schwinger-type integral equation, which is subsequently resolved through iterative techniques.

One of the primary advantages of FFT-based homogenization is its mesh-free nature, utilizing a uniform grid that significantly simplifies preprocessing steps, especially for intricate microstructures. This attribute renders the method exceptionally compatible with high-resolution imaging data, such as X-ray tomography or image-based microstructures. Additionally, the inherent computational efficiency of FFT-based methods, when combined with rapid Fourier transforms, results in reduced memory requirements compared to traditional Finite Element Methods (FEM), making them highly suitable for large-scale simulations.

At the mesoscopic level, pioneering contributions by Brenner have been instrumental in adapting FFT-based methods to electromechanical composites. Brenner \cite{Brenner2009} was among the first to incorporate piezoelectricity into the FFT framework, implementing the foundational strain-based iterative scheme initially proposed by Moulinec and Suquet in two dimensions. To address convergence challenges in scenarios involving high phase contrast and uncoupled problems, Brenner further developed the augmented Lagrangian FFT approach. This advancement facilitated the creation of iterative schemes capable of handling coupled piezoelectric and magneto-electro-elastic constitutive relations, as demonstrated in subsequent works by Brenner \cite{Brenner2010} and Brenner and Bravo-Castillero \cite{BrennerBC2010}. Notably, these implementations have predominantly been confined to two-dimensional analyses.

Recent advancements have extended the applicability of spectral methods to the microscopic level in ferroelectric materials, usually in combination with phase-field methods, as described in the following.

\subsubsection{Phase-field method for domain evolution in ferroelectric materials}

The phase-field method is a powerful computational approach for modeling the microstructural evolution of ferroelectric materials, particularly in capturing complex phenomena such as domain formation, domain wall motion, and phase transitions. Building upon the electromechanical and thermo-electromechanical models presented in Sections 4.1 to 4.7, the phase-field method introduces spatially and temporally resolved polarization fields, enabling detailed simulations of domain dynamics under varying external stimuli.

At the core of the phase-field method lies the time-dependent Ginzburg-Landau (TDGL) equation, already introduced in Eq.~(77), which governs the temporal evolution of the polarization vector $\overrightarrow{p}$. This equation is derived from the minimization of the total free energy of the system, as defined in Eq. (25). The TDGL equation facilitates the simulation of polarization dynamics by accounting for both local and non-local interactions within the material. The total free energy functional $\phi$ encompasses contributions from the Landau-Ginzburg-Devonshire (LGD) free energy $W(\theta, \overrightarrow{p})$, gradient energy associated with spatial variations in polarization, elastic energy due to mechanical strains, and electrostatic energy from the interaction between electric fields and polarization. Specifically, the gradient energy term $\frac{1}{2} \lambda |\nabla \overrightarrow{p}|^2$ promotes the formation of domain walls by penalizing sharp polarization gradients.

The phase-field method integrates seamlessly with the homogenization framework discussed in the previous section and can be effectively coupled with spectral techniques to harness the computational efficiency of Fourier transforms. This integration enables detailed microscale resolution of polarization and strain fields, providing insights into how microstructural features impact the macroscopic effective properties of the material. Such modeling precision is crucial for designing lead-free piezoelectric materials with tailored properties for specific applications. 

To simulate domain evolution, the phase-field method discretizes the governing equations spatially using numerical techniques such as finite difference, finite element, or spectral methods, as detailed in Section 4.9. Specifically, in combination with spectral techniques, the phase-field method facilitates efficient resolution of coupled elasto-electrostatic equilibrium equations in the transformed domain. 

The recent contribution by Durdiev and Wendler \cite{DURDIEV2023111928} exemplifies the synergy between phase-field modeling and spectral methods. Their Fourier spectral phase-field approach adeptly simulates domain formation and polarization switching in tetragonal perovskite ferroelectrics under varying electric fields and mechanical stresses. By incorporating spontaneous polarization as the primary order parameter and integrating piezoelectric effects into the constitutive framework, their model accurately reproduces domain evolution phenomena observed experimentally.

Further refinements in spectral phase-field methods for ferroelectric ceramics have been achieved by Kochmann and colleagues \cite{VIDYASAGAR2017133, INDERGAND2023105426}. They developed an enhanced spectral phase-field method capable of predicting the effective electromechanical responses of bulk polycrystalline ferroelectrics, such as BaTiO$_3$ and Pb(Zr,Ti)O$_3$ (PZT). Their approach effectively models complex polycrystalline microstructures, providing insights into electric hysteresis and strain-electric field responses. Additionally, they addressed numerical challenges like instability and ringing phenomena through finite-difference approximations. They propose to assume homogeneous microstructure so no iterative solution scheme is required.  Indergand et al. \cite{INDERGAND2023105426} further advanced this field by conducting high-resolution simulations of ferroelectric domain evolution in PZT, utilizing a Fourier spectral scheme to achieve atomic-scale resolution. Their work underscores the significant influence of grain orientation on domain patterns, linking monodomain and laminate structures to crystallographic orientations and correlating these configurations with macroscopic piezoelectric and dielectric properties validated through high-energy X-ray diffraction experiments.

In a recent study by Akshayveer et al., \citep{Akshayveer2024a}, the influence of ferroelectric domain switching and phase change behaviour of the BNT-PDMS composite is examined utilizing the phase-field approach. A two-dimensional computational model has been created to examine the impact of phase-field thermoelectromechanical performance on the lead-free BNT-PDMS composite under diverse temperature and haptic stress circumstances. The complex phase and domain structures, entirely reliant on the material's temperature, become increasingly convoluted at elevated temperatures due to the temporal evolution of spontaneous polarization; hence, periodic boundary conditions are essential as domain walls swap arbitrarily. Nonetheless, BNT has complicated phase shift behaviour, but the PDMS matrix remains unresponsive to such alterations; hence, accurate assessment of effective characteristics is essential, given that BNT inclusions are distributed randomly or in a specific manner inside the PDMS matrix. The phase-field model accurately simulates the phase transition and micro-dynamic behaviour of BNT-based composites, with the anticipated effective piezoelectric coefficients closely aligning with experimental values for BNT \citep{Hiruma2009}. Additionally, we have investigated the phase change behaviour of pure BNT utilizing a phase-field model, without employing effective property estimates \citep{Akshayveer2024}. The phase-field technique demonstrates P-E curves analogous to those reported in experimental investigations of BNT \citep{WANG20164268}. Therefore, it can be stated that the phase-field technique is an effective approach to investigate and comprehend the intricate phase change behaviour of ferroelectric materials.

In summary, the phase-field method extends the theoretical models established in previous sections by providing a robust framework for simulating the dynamic evolution of polarization domains. This enhanced modeling capability is instrumental in understanding and optimizing the performance of lead-free piezoelectric materials in various functional applications.

Apart from the solution scheme, there can be other methods such as artificial intelligence (AI) generated machnie learning (ML) tools to enhance the efficacy of the numerical models by making them more economic and time-efficient. A synopsis about applied ML has been provided in next subsection.

\subsection{Applied Machine Learning for Developing Next-Generation Functional Materials}

Machine learning (ML) is transforming the identification and enhancement of functional materials, particularly in the creation of lead-free piezoelectric materials. Conventional experimental techniques for material discovery and property enhancement are frequently laborious and resource-demanding. Utilizing extensive datasets and predictive models, machine learning expedites the development of materials with specific features, facilitating swift progress in sustainable piezoelectric materials. Recent research, like those presented in \citep{Dinic2021, SINGH2024}, underscore the substantial impact of machine learning in the field of materials science.

\subsubsection{Data-Driven Materials Discovery}

In materials science, machine learning models are utilized to forecast essential material characteristics, including piezoelectric coefficients, thermal stability, and mechanical durability. These models depend on extensive databases of material attributes, which are analyzed to reveal relationships among composition, structure, and functioning. In lead-free piezoelectrics, machine learning approaches such as regression analysis \citep{QI2019}, deep learning \citep{GU2018}, and Bayesian optimization \citep{MartinezCantin2017} are employed to forecast material behaviour based on atomic composition and crystallographic characteristics. Researchers have created machine learning models that forecast the performance of bismuth sodium titanate (BNT)-based materials by examining trends in current experimental data \citep{Sun2024}. These models assist in determining ideal dopants and compositions that enhance piezoelectric performance while maintaining environmental sustainability \citep{Jing2023}.

\subsubsection{Accelerating the Design of Lead-Free Piezoelectrics}

The capacity of machine learning to rapidly assess prospective materials has been essential for the advancement of lead-free substitutes for conventional lead zirconate titanate (PZT) systems. Researchers can substantially decrease the number of experimental trials needed to evaluate novel compositions by employing algorithms that replicate material behaviour. In this regard, generative models have been utilized to develop innovative lead-free piezoelectric materials, such sodium potassium niobate (KNN) \citep{SAPKAL2023} and barium titanate (BaTiO$_3$) \citep{YUAN20221} composites with improved characteristics.

A significant improvement is employing random forest and neural networks to predict piezoelectric responses in intricate material systems. These approaches have facilitated the forecasting of high-performance compositions that are more sustainable, directing experimental endeavours toward potential candidates.

\subsubsection{Materials Optimization through Active Learning}

Active learning, a method in materials science, is one of the most promising uses of machine learning, wherein the model engages with experimental outcomes to progressively enhance its predictions \citep{Lookman2019,Huang20231}. Active learning can enhance the piezoelectric capabilities of lead-free materials by perpetually updating models informed by real-time experimental input. This method minimizes the quantity of tests required to get optimal material performance, hence accelerating the development cycle considerably.

\subsubsection{Predictive Modeling for Device Integration}

Machine learning is utilized to simulate the behaviour of piezoelectric materials in practical applications, including sensors and actuators \citep{Hu2022,Ferreira2024,Castillo2021}. Machine learning aids in customizing materials for particular applications by modelling their behaviour under various operating situations, including micro-electromechanical systems (MEMS), biomedical devices, and energy harvesters. Models developed using experimental data have effectively predicted the behaviour of lead-free piezoelectrics under diverse mechanical loads and temperatures, facilitating the advancement of durable materials for next-generation technology.

\vspace{10px} 

The use of machine learning in the advancement of lead-free piezoelectric materials signifies a crucial progression towards the creation of more sustainable and high-performance functional materials. Machine learning is crucial in accelerating the discovery, optimization, and implementation of sustainable materials that satisfy contemporary technological requirements. Ongoing progress in data-driven materials science is anticipated to augment the performance of lead-free piezoelectrics, facilitating their extensive use across diverse industrial and technological sectors.

\vspace{10px}
The preceding section comprises a compilation of several mathematical models and solution methodologies, which may be advantageous for applications involving devices utilizing lead-free piezoelectric materials. The choice of a mathematical model and solution method is contingent upon the reliability of the intended outcome about input variables. Occasionally, the electromechanical model and RVE technique suffice for a low number of input variables due to the constraints of intended output and ambient circumstances in the scenario. Furthermore, the issue may be as intricate as ferroelectric polarization, which fluctuates at sub-micrometer scales, hence augmenting the input variables necessary to get the desired output for device applications. Consequently, the spectrum solution scheme may serve as a feasible choice for addressing such problems in a time-efficient and cost-effective manner. This research not only presents information on the utility of various lead-free materials for certain applications but also offers mathematical models and solution schemes that facilitate the analysis of their performance in a time-efficient and cost-effective manner which can further be accelerated by using applied machine learning. Furthermore, a synopsis of the complete paper is included in the subsequent section.

\section{Summary}
\markright{Summary \hfill \empty}

The improvements and uses of environmentally friendly technologies that make use of lead-free piezoelectric materials are taken into consideration in this review study. Lead-free alternatives have become increasingly popular among academics as worries about the environmental effect of lead-based materials have risen. These alternatives not only reduce the hazards to the environment, but they also display competitive performance in a variety of applications.
High performance lead-free piezoelectric ceramics have been the spotlight
of in recent years due to increased demands of eco-friendly energy
harvesting techniques. There is a complete summary of lead-free piezoelectric materials presented at the beginning of the study. These materials include barium titanate (BaTiO$_{2}$), sodium potassium niobate (KNN), bismuth sodium titanate (BNT), bismuth layerered structured ferroelectrics (BLSF), Li-based piezoelectrics and many others such as the composites of above discussed materials. This article discusses the synthesis, properties, and performance characteristics of various materials, with a particular emphasis on the piezoelectric responses, thermal stability, and mechanical properties for these materials. 

Following this introduction, the paper digs into the many uses of lead-free piezoelectric materials across a variety of industries, such as instruments for energy harvesting, sensing, actuating, biomedical and resonating devices. A particular emphasis is focused on the development of new technologies for energy harvesting, such as vibration energy harvesters, wearable devices, and structural health monitoring systems that make use of lead-free piezoelectric materials to transform mechanical energy into electrical energy in a manner that is environmentally friendly. The incorporation of these materials into sensor technologies and actuators further highlights the adaptability and significance of these materials in the creation of intelligent devices and systems. This research further examines alternative eco-friendly applications, including haptic devices, nano-electromechanical devices, photostrictive devices, and piezo-photostrictive devices. Lead-free materials have significant potential for optimal performance in these applications.

As an additional point of interest, the paper investigates current advancements in nanostructured and composite piezoelectric materials. These materials improve the performance of lead-free systems and widen their utility in advanced technologies. In addition, the topic touches on the difficulties and constraints that are associated with lead-free materials. These include the fact that lead-free materials have lower piezoelectric coefficients in comparison to their lead-containing counterparts, as well as the continuous research efforts that are being made to optimize their characteristics for certain applications.

However, the majority of the research on lead-free piezoelectric materials has been carried out through experimental methods, which is a procedure that is both time-consuming and expensive. Despite this, the study is still ongoing. Despite the fact that experimental discoveries serve as the foundation for computational studies, computational studies are an approach that is both cost-effective and efficient for determining the way in which these materials operate in a variety of applications. In the latter portion of this article, the emphasis is placed on delivering state-of-the-art theoretical and modelling techniques under a variety of applied external force and energy fields in order to undertake advanced research in a manner that is both more expedient and less expensive. The macroscopic modelling and the micro-domain modelling for phase change, domain switching, and viscoelastic damping of lead-free piezoelectric materials are the primary areas of focus for the numerical models that are being studied in this article. This paper looked at a number of different solution schemes for solving the partial differential equations of the mathematical models that were discussed earlier. These solution schemes include the FDM, FVM, FEM, RVE methods, and the spectral solution scheme. In addition to the theoretical models, this paper also included other solutions. This study report not only describes the technique for finding a solution, but it also discusses the benefits and drawbacks of the numerous different schemes, as well as how well they fit with the mathematical models that were mentioned before. In addition, the use of piezoelectric material and the output that is wanted from the device were taken into consideration while selecting the mathematical model and the solution method. In order to get the best possible results in terms of both cost and time efficiency, the mathematical model and solution scheme are integrated with ML tools to get it  done in more time-efficient and economic way. An important step forward in this field is the integration of high-throughput computational tools and curated databases, as demonstrated by the Materials Project’s expanded dataset for piezoelectric materials, which was meticulously analyzed in this study to unveil new opportunities for lead-free compounds.

Moreover, choosing the appropriate lead-free piezoelectric material for a certain application and selecting the appropriate mathematical model and solution scheme to examine the performance of that specific piezoelectric devices in the most cost-effective and time-efficient manner are the two factors that are the primary emphasis of this paper. So, not only will this review article be useful in the investigation of lead-free piezoelectric materials for the purpose of various enivironmentally friendly applications (such as energy harvesting devices, haptic devices, nano-electromechanical devices, photostrictive devices, piezo-photostrictive devices, sensors and actuators for biomedical devices) but it will also give us the most computationally efficient modelling methodologies that will allow us to carry out advanced study in this field.

\section*{Acknowledgements}
\markright{Acknowledgements \hfill \empty}

The authors are grateful to the NSERC and the CRC Program (Canada) for their support. This publication is part of the R+D+i project, PID2022-137903OB-I00, funded by MI-CIU/AEI/10.13039 /501100011033/ and by ERDF/ EU. This research was enabled in part by support provided by SHARCNET (www.sharcnet.ca) and Digital Research Alliance of Canada (www.alliancecan.ca).

%
\markright{References \hfill \empty}
\bibliographystyle{elsarticle-num}
\addcontentsline{toc}{section}{\refname}\bibliography{Environment-friendly_References}

\section*{\textemdash \textemdash \textemdash \textemdash \textemdash \textemdash \textemdash{}}
\end{document}